\documentclass[11pt]{article}
\usepackage{amsmath,latexsym,fullpage,color, subfigure, rotating,url,setspace, multirow, amssymb, authblk}
\title{Efficient Algorithms for Multivariate Linear Mixed Models in Genome-wide Association Studies}
\author[1]{Xiang Zhou \thanks{xz7@uchicago.edu}}
\author[1,2]{Matthew Stephens \thanks{mstephens@uchicago.edu}}
\affil[1]{Department of Human Genetics, University of Chicago}
\affil[2]{Department of Statistics, University of Chicago}
\date{}

\newcommand{\bb}{\mathbf{b}}
\newcommand{\be}{\mathbf{e}}
\newcommand{\bg}{\mathbf{g}}

\newcommand{\bs}{\mathbf{s}}
\newcommand{\bq}{\mathbf{q}}

\newcommand{\bx}{\mathbf{x}}
\newcommand{\by}{\mathbf{y}}

\newcommand{\bw}{\mathbf{w}}

\newcommand{\bK}{\mathbf{K}}
\newcommand{\bV}{\mathbf{V}}
\newcommand{\bA}{\mathbf{A}}
\newcommand{\bB}{\mathbf{B}}
\newcommand{\bX}{\mathbf{X}}
\newcommand{\bY}{\mathbf{Y}}
\newcommand{\bE}{\mathbf{E}}
\newcommand{\bG}{\mathbf{G}}
\newcommand{\bH}{\mathbf{H}}
\newcommand{\bP}{\mathbf{P}}
\newcommand{\bQ}{\mathbf{Q}}
\newcommand{\bR}{\mathbf{R}}
\newcommand{\bW}{\mathbf{W}}

\newcommand{\bU}{\mathbf{U}}

\newcommand{\bD}{\mathbf{D}}
\newcommand{\bI}{\mathbf{I}}

\newcommand{\bbeta}{\boldsymbol\beta}

\newcommand{\btheta}{\boldsymbol\theta}

\newcommand{\bSigma}{\boldsymbol\Sigma}
\newcommand{\bLambda}{\boldsymbol\Lambda}

\begin{document}
\maketitle

\begin{abstract}
Multivariate linear mixed models (mvLMMs) have been widely used in many areas of genetics, and have attracted considerable recent interest in genome-wide association studies (GWASs). 
However, fitting mvLMMs is computationally non-trivial, and no existing method is computationally practical for 
performing the likelihood ratio test (LRT) for mvLMMs in GWAS settings with moderate sample size $n$. The existing software MTMM \cite{Korte:2012} perform an approximate LRT for two phenotypes, and as we find, its $p$ values can substantially understate the significance of associations. Here, we present novel computationally-efficient algorithms for fitting mvLMMs, and computing the LRT in GWAS settings. After a single initial eigen-decomposition (with complexity $O(n^3)$) the algorithms  i) reduce computational complexity (per iteration of the optimizer) from cubic to linear in $n$; and ii) in GWAS analyses, reduces per-marker complexity from cubic to quadratic in $n$. These innovations make it practical to compute the LRT for mvLMMs in GWASs for tens of thousands of samples and a moderate number of phenotypes ($\sim2-10$). With simulations, we show that the LRT provides correct control for type I error. With both simulations and real data we find that the LRT is more powerful than the approximate LRT from MTMM, and illustrate the benefits of analyzing more than two phenotypes. The method is implemented in the GEMMA software package, freely available at \url{http://stephenslab.uchicago.edu/software.html}.
\end{abstract}

\section*{Introduction}
Multivariate linear mixed models (mvLMMs) \cite{Henderson:1984} have been applied to many areas of genetics, including, for example, estimating the cross-tissue heritability of gene expression \cite{Price:2011}, assessing the pleiotropy and genetic correlation between complex phenotypes \cite{Korte:2012, Lee:2012b, Vattikuti:2012, Trzaskowski:2013}, detecting quantitative trait loci \cite{Amos:1994}, understanding evolutionary patterns \cite{Kruuk:2004} and assisting animal breeding programs \cite{Meyer:2004}. Recently, these models have become increasingly important in genome-wide association studies (GWASs), not only because of their demonstrated effectiveness in accounting for sample relatedness \cite{Meyer:1991, Amos:1994, Korte:2012} and in controlling for population stratification \cite{Korte:2012} (as for their univariate counterparts \cite{Abney:2002, Yu:2006, Chen:2007, Kang:2008, Zhang:2010, Kang:2010, Lippert:2011, Zhou:2012, Pirinen:2013, Svishcheva:2012, Zhou:2013}), but also because of a growing appreciation of the potential gains in power from multivariate association analyses \cite{Banerjee:2008, Kim:2009a, Kim:2009b, Ferreira:2009, Zhang:2009, OReilly:2012, Korte:2012, Stephens:2013}.  Indeed,
\cite{Stephens:2013} emphasizes that, compared with standard univariate
analyses, multivariate analyses can increase power not only to detect pleiotropic genetic variants that affect multiple phenotypes simultaneously, but also genetic variants that affect only one of a collection of correlated phenotypes.

However, fitting mvLMMs is computationally non-trivial, involving a multi-dimensional optimization for a potentially non-convex function (either the likelihood or the restricted likelihood). These computational challenges 
become very substantial
when applying mvLMMs to GWASs: for example, performing the likelihood ratio test (LRT) for association with each SNP requires these optimizations to be performed repeatedly for hundreds of thousands of SNPs in turn.
Consequently no existing method is computationally practical for 
performing the LRT for mvLMMs in GWAS settings. The only available
method along these lines (MTMM \cite{Korte:2012}) can perform only an approximate LRT, and only for two phenotypes. Furthermore, as we will
show later, the $p$ values from this approximation are not well calibrated under the null hypothesis, at least in settings involving strongly related individuals, and can substantially understate the significance of associations.

 Here, we present a novel computationally-efficient algorithm, and a software implementation, for fitting mvLMMs with one covariance component (in addition to the residual error term), and for performing the LRT of association in GWASs. The algorithm builds on linear algebra techniques previously used for univariate LMMs \cite{Lippert:2011, Zhou:2012, Pirinen:2013}, and, combined with
 several additional tricks, 
extends them to multivariate LMMs.  In effect, our algorithms provide the multivariate analogues of the univariate algorithms EMMA \cite{Kang:2008} and  FaSTLMM/GEMMA/CM \cite{Lippert:2011, Zhou:2012, Pirinen:2013}. Our algorithms provide the first computationally-practical approach to computing LRTs for mvLMMs in GWAS with a reasonably large number of individuals (e.g.~50,000), and for a modest  number of phenotypes (e.g.~2-10).

\section*{Results}
\subsection*{Model and method overview}
We consider the multivariate linear mixed model \cite{Henderson:1984},
\begin{equation}
\bY=\bA\bW+\bbeta\bx^T+\bG+\bE; \quad \bG \sim \mbox{MN}_{d \times n}({\bf 0}, \bV_g, \bK), \quad \bE \sim \mbox{MN}_{d \times n}({\bf 0}, \bV_e, \bI_{n\times n}),
\end{equation}
where $n$ is the number of individuals, $d$ is the number of phenotypes, $\bY$ is a $d$ by $n$ matrix of phenotypes, $\bW$ is a $c$ by $n$ matrix of covariates including a row of $1$s as intercept and $\bA$ is a $d$ by $c$ matrix of corresponding coefficients, $\bx$ is a $n$-vector of genotypes for a particular marker and $\bbeta$ is a $d$-vector of its effect sizes for the $d$ phenotypes, $\bG$ is a $d$ by $n$ matrix of random effects, $\bE$ is a $d$ by $n$ matrix of residual errors, $\bK$ is a known $n$ by $n$ relatedness matrix, $\bI_{n\times n}$ is the $n$ by $n$ identity matrix, $\bV_g$ is a $d$ by $d$  symmetric matrix of genetic variance component, $\bV_e$ is a $d$ by $d$  symmetric matrix of environmental variance component and $\mbox{MN}_{d \times n}(\bf{0}, \bV_1, \bV_2)$ denotes the $d \times n$ matrix normal distribution with mean 0, row covariance matrix $\bV_1$ ($d$ by $d$), and column covariance matrix $\bV_2$ ($n$ by $n$). 

We have developed efficient algorithms for applying mvLMMs to GWASs. Specifically, for each genetic marker in turn, the algorithms perform the LRT comparing the null hypothesis that the marker effect sizes for all phenotypes are zero, $H_0: \bbeta= \bf 0$, where $\bf 0$ is a $d$-vector of zeros, against the general alternative $H_1: \bbeta\neq \bf 0$. 
Our algorithms provide the first computationally-practical approach to computing this LRT for GWAS, and our software
is the only available implementation for computing mvLMM test statistics of any kind  in GWAS with more than 2 phenotypes.

The LRT requires maximum likelihood estimates 
for parameters $\bV_g, \bV_e, \bA$ and $\bbeta$ under both $H_0$ and $H_1$.  Current algorithms for obtaining these estimates (implemented in software GCTA \cite{Yang:2011b, Lee:2012b}, WOMBAT \cite{Meyer:2007}, ASREML \cite{Gilmour:1995}) all use similar approaches with the same computational complexity. Specifically, the methods combine two types of optimization algorithm: an initial Expectation-Maximisation-like (EM) algorithm, followed by a Newton-Raphson-like (NR) algorithm. This combines the benefits of the
stability of the EM-like algorithm (every iteration increases the likelihood) 
with the faster convergence of the NR-like algorithm (\cite{Meyer:2006}; Supplementary Note and Figure \ref{sfig:pxemvsboth}). 
The computational complexity of these existing methods is $O(t_1n^3d^3+t_2n^3d^7)$ where
$t_1,t_2$ are the maximum number of iterations used for EM and NR optimizers respectively (Supplementary Note). To apply these to GWAS would require this procedure to be repeated for all $s$ SNPs, with resulting computational complexity $O(s(t_1n^3d^3+t_2n^3d^7))$. This is computationally impractical for GWAS with large $s$ and moderate $n$ (see below).

In comparison, our algorithms very substantially reduce the computational burden of computing LRTs for GWAS, by using linear algebra tricks to avoid repeating the expensive $O(n^3)$ operations for every SNP. 
Specifically, after an initial single $O(n^3)$ operation (eigen-decomposition of the relatedness matrix $K$), our algorithms have per-SNP complexity that increases only quadratically in $n$.  The overall computational complexity is $O(n^3+n^2d+s(n^2+t_1nd^2+t_2nd^6))$. In effect, our algorithms (detailed in Supplementary Note)
provide the multivariate analogue of the univariate algorithms
EMMA \cite{Kang:2008}, and FaSTLMM/GEMMA/CM \cite{Lippert:2011, Zhou:2012, Pirinen:2013}. 

We implemented the algorithms in the GEMMA software package \cite{Zhou:2012, Zhou:2013}, freely available at \url{http://stephenslab.uchicago.edu/software.html}.

\subsection*{Comparisons with existing methods}

To illustrate the benefits of our new algorithms we used two data sets: a mouse GWAS from the Hybrid Mouse Diversity Panel (HMDP) with four blood lipid phenotypes (high-density lipoprotein, HDL; total cholesterol, TC; triglycerides, TG; unesterified cholesterol, UC), and a human GWAS from the Northern Finland Birth Cohort 1966 (NFBC1966) with four blood metabolic traits (high-density lipoprotein, HDL; low-density lipoprotein, LDL; triglycerides, TG; C-reactive protein, CRP). The HMDP data are a small GWAS ($n=656$) with strong relatedness among many individuals; the NFBC1966 data are a larger GWAS ($n=5,255$) with weak relatedness
among most individuals.

Even for fitting a single mvLMM, our algorithms have lower computational complexity than existing algorithms. Indeed, for these data
our implementation in GEMMA is substantially (5-71 times) faster than implementations of existing algorithms in software GCTA and WOMBAT (Table \ref{tab:time_null}). For example, for the NFBC1966 data, with $d=4$, GEMMA takes about 7 minutes compared with 8 hours for WOMBAT. The gains for larger $d$ would be even greater. 

\begin{table}
\begin{center}
\caption{Comparison of computation time of different methods for parameter estimation in a single mvLMM. Results are shown for both HMDP and NFBC1966 data sets. All computing was performed on a single core of an Intel Xeon L5420 2.50GHz CPU. In all cases the three packages give numerically identical results, both for estimated variance components and their standard errors (data not shown). $n$ is the number of individuals, $d$ is the number of traits, $c$ is the number of covariates ($c=1$ here), $t_1$ is the number of iterations used in the EM type algorithm and $t_2$ is the number of iterations used in the NR type algorithm. The implementation of GCTA does not handle more than two traits. The comparison also favors the other two software because of the much more stringent stopping criteria used in GEMMA: $t_1$ in GEMMA is often two orders of magnitude larger than that used in the other two software, while $t_2$ in GEMMA is similar to that used in the other two. The $n^3$ step in GEMMA could be replaced with a $mn^2$ step \cite{Kang:2008} for the HMDP data, where $m$ is the number of strains. Notice that the compute time for GEMMA is essentially the same for all $d$, because in GEMMA the compute time is dominated by the initial $O(n^3)$ eigen-decomposition step; the following optimization iterations are negligible. }
\label{tab:time_null}
\bigskip
\begin{small}
\begin{tabular}{ cccccccc }
\hline
\multirow{4}{*}{Method}  & \multirow{4}{*}{Time Complexity} & \multicolumn{6}{c}{Computation Time} \\
&&\multicolumn{3}{c}{HMDP} & \multicolumn{3}{c}{NFBC1966 } \\
&&\multicolumn{3}{c}{($n=656$)} & \multicolumn{3}{c}{($n=5255$)} \\
& & $d=2$ & $d=3$ & $d=4$ &$d=2$ &$d=3$ &$d=4$ \\
\hline
GEMMA & $O(n^3+n^2d+n^2c+t_1nc^2d^2+t_2nc^2d^6)$ & $<1.0$ s & $<1.0$ s & $<1.0$ s & 6.7 min & 6.7 min & 6.7 min\\
WOMBAT & $O(t_1n^3(d+c)^3+t_2n^3d^7)$ & 12.5 s & 39.2 s & 71.0 s & 31.0 min & 127.6 min & 477.3 min\\
GCTA & $O(t_1n^3(d+c)^3+t_2n^3d^7)$ & 11.2 s & -- & -- & 38.2 min & -- & --\\
\hline
\end{tabular}
\end{small}
\end{center}
\end{table}

However, the more practically-important gains of our new algorithms come in GWAS applications.
Here, no existing algorithm is practical for computing the LRT for even $d=2$. Extrapolating from Table \ref{tab:time_null}
suggests that existing algorithms, if implemented in software, might take over 14 days for HMDP and over 18 years 
for NFBC1966. As far as we are aware, the only current practical competitor for our method is a method implemented in software MTMM
 \cite{Korte:2012}, which uses an approximation to the LRT  \cite{Aulchenko:2007, Zhang:2010, Kang:2010} to reduce per-SNP computation time to quadratic in $n$.  Specifically, this approximate LRT avoids the expensive repeated optimization of the variance components under $H_1$ for each SNP, by re-using part of the pre-estimated variance components under $H_0$ (fit using the software ASREML).
However, this approximate LRT is guaranteed to underestimate the LRT (Supplementary Note), and in the univariate setting this has been shown to produce mis-calibrated $p$ values and/or loss of power for data sets involving smaller numbers of individuals and
strong relatedness \cite{Zhou:2012, Pirinen:2013}.

\begin{figure}[htb!]
\centering
 \subfigure[HMDP-based simulations, Type I Error]{
   \includegraphics[width=210pt, height=210pt, angle=0]{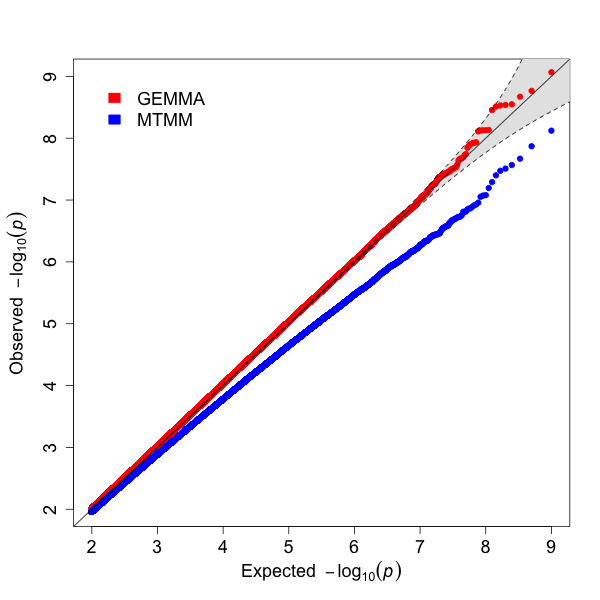}
   \label{sfig:t1e}
 }
 \subfigure[HMDP-based simulations, Power]{
   \includegraphics[width=210pt, height=210pt, angle=0]{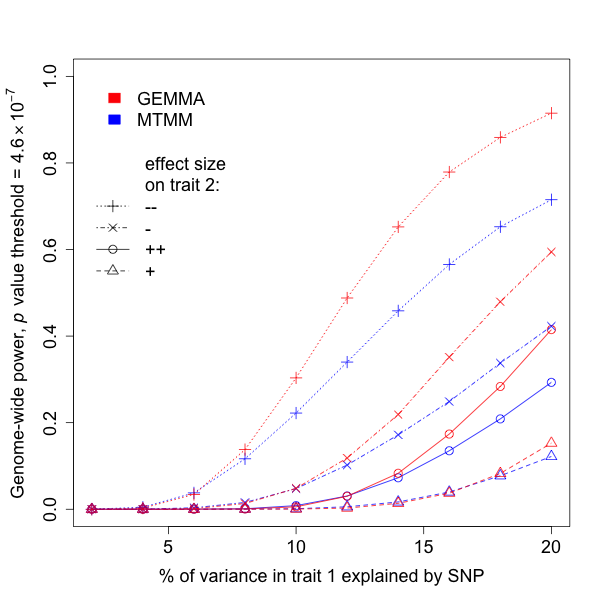}
   \label{sfig:power}
 }
 \subfigure[HMDP]{
   \includegraphics[width=210pt, height=210pt, angle=0]{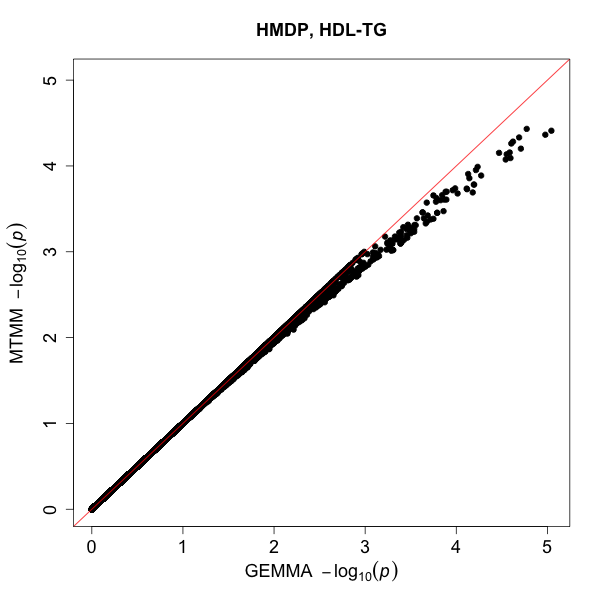}
   \label{fig1:a}
 }
 \subfigure[NFBC1966]{
   \includegraphics[width=210pt, height=210pt, angle=0]{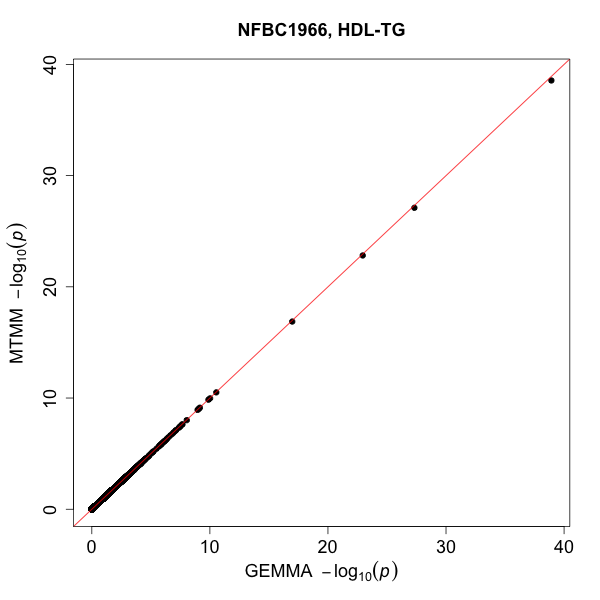}
   \label{fig1:b}
 }
\caption{GEMMA versus MTMM. (a) Comparison of type I error control with simulations based on the HMDP data. QQ-plot of observed versus expected -$\log_{10}p$ values under the null, for GEMMA (red) and MTMM (blue). Gray shaded area indicates 0.025 and 0.975 point-wise quantiles of the ordered $p$ values under the null distribution. (b) Comparison of power for GEMMA (red) and MTMM (blue), at a nominal $p$ value threshold of $4.6 \times 10^{-7}$, in four different simulation scenarios based on the HMDP data. x-axis in shows the proportion of phenotypic variance in the first phenotype explained (PVE) by the SNP, while the point symbol and line type indicate the SNP effect direction (compared with its effect on the first phenotype) and size (quantified by PVE) on the second phenotype (+: opposite direction, 0.8PVE; $\times$: opposite direction, 0.2PVE; o: same direction, 0.8PVE; $\Delta$: same direction, 0.2PVE). Comparison of -$\log_{10}p$ values for (c) paired traits HDL-TG  in the HMDP data and (d) paired traits HDL-TG in the NFBC1966 data. }
\label{fig:twotrait_pval}
\end{figure}

 To illustrate this in the multivariate setting we performed null and alternative simulations
using the HMDP data (Online Methods). Consistent with the univariate findings, MTMM $p$ values are systematically larger
than expected under the null, with the most significant $p$ values
being almost an order of magnitude larger than expected (Figure \ref{fig:twotrait_pval}a).
In contrast, $p$ values from the GEMMA LRT are well-calibrated (Figure \ref{fig:twotrait_pval}a). 
This demonstrates that, despite the fact that the mvLMM likelihood surface could be non-convex, with multiple local optima, any problems that
our optimization methods have with local maxima have minimal practical impact. (We found that obtaining 
well-calibrated $p$ values requires both the EM and NR algorithms: use of only the EM algorithm can 
lead to poor convergence of the LRT, and resulting in underestimation of $p$ values similar to MTMM; Figure \ref{sfig:pxemvsboth})
The systematic inflation of MTMM's $p$ values under the null presumably accounts for MTMM's loss of power relative to GEMMA in simulations under the alternative (Figure \ref{fig:twotrait_pval}b).

In addition to these simulations, we also compared GEMMA and MTMM on the real
HMDP and NFBC1966 datasets.
Since MTMM is implemented only for $d=2$, we analyzed all pairs of traits. For these data, GEMMA ran 2-12 times faster than MTMM (Table \ref{tab:time_gwas}).  In particular, in the NFBC1966 data, GEMMA takes about four hours to finish a two-phenotype mvLMM analysis that takes MTMM almost two and a half days. (In fact, GEMMA can finish the multivariate analysis for four traits within six hours.)
Consistent with the simulations, and with theory, the MTMM $p$ values for HMDP are consistently less significant (up to 6 fold less significant) than $p$ values from GEMMA (Figure \ref{fig:twotrait_pval}c). For NFBC1966 the two methods produce similar $p$ values (Figure \ref{fig:twotrait_pval}d), consistent with 
univariate assessments that show the approximation used in MTMM to work well when, as in NFBC1966, the sample size is large, individuals are not closely related and the marker effect size is small.

\begin{table}
\begin{center}
\caption{Computation time for GWAS analysis with mvLMM using different methods for the HMDP and NFBC1966 data sets. All computing was performed on a single core of an Intel Xeon L5420 2.50GHz CPU. The computing time for MTMM only include the multiplication of the genotype matrix with an $nd$ by $nd$ matrix, and does not include the time spent fitting the null model (because MTMM relies on the commercial software ASREML to do so), nor the time spent reading/writing files and inverting the $nd$ by $nd$ matrix. The computing time for GEMMA include all steps. $n$ is the number of individuals, $s$ is the number of SNPs, $d$ is the number of traits, $c$ is the number of covariates ($c=1$ here), $t_1$ is the number of iterations used in the EM type algorithm and $t_2$ is the number of iterations used in the NR type algorithm. The MTMM software implementation handles only two traits. The $pn^2$ step in GEMMA could be replaced with a $pnr$ step if the relatedness matrix is of rank $r$.}
\label{tab:time_gwas}
\bigskip
\begin{small}
\begin{tabular}{ cccccccc }
\hline
\multirow{4}{*}{Method}  & \multirow{4}{*}{Time Complexity} & \multicolumn{6}{c}{Computation Time} \\
& &\multicolumn{3}{c}{HMDP} & \multicolumn{3}{c}{NFBC1966 } \\
& &\multicolumn{3}{c}{($n=656$, $s=108,562$)} & \multicolumn{3}{c}{($n=5255$, $s=319,111$)} \\
& & $d=2$ & $d=3$ & $d=4$ & $d=2$ & $d=3$ & $d=4$ \\
\hline
GEMMA & $O(n^3+n^2d+n^2c+s(n^2+t_1nc^2d^2+t_2nc^2d^6))$ & 6.2 min & 13.7 min & 28.5 min & 4.4 h & 4.8 h & 5.8 h\\
MTMM & $O(t_1n^3(d+c)^3+t_2n^3d^7+sn^2d^2)$ & 16.4 min & -- & -- & 58.0 h & -- & --\\
\hline
\end{tabular}
\end{small}
\end{center}
\end{table}

\subsection*{Analyzing more than two phenotypes}
Our methods and software also make possible, for the first time, GWAS analysis using mvLMMs with more than two phenotypes.
The use of multivariate association analyses in GWAS has 
been the subject of considerable recent interest \cite{Banerjee:2008, Kim:2009a, Kim:2009b, Ferreira:2009, Zhang:2009, OReilly:2012, Korte:2012, Stephens:2013}, and
many studies have demonstrated the potential gain in power of multivariate vs univariate analyses \cite{Banerjee:2008, Kim:2009a, Kim:2009b, Ferreira:2009, Zhang:2009, OReilly:2012, Korte:2012, Stephens:2013}. However, multivariate analyses also raise their own  challenges, not least that a significant result in a multivariate test does not immediately indicate which phenotypes are driving the association. These issues are discussed in detail in \cite{Stephens:2013},
which outlines a framework for multivariate association analysis based
on model comparison, rather than testing.  This framework also clarifies
under which situations various multivariate tests may be more or less powerful
than univariate tests, and provides guidance on how to integrate results
from multiple association tests (e.g. from both univariate and multivariate tests).  Our work here lays the foundations for using this framework
with mvLMMs; specifically, our maximum likelihood algorithms could be used to approximate (e.g. via Laplace approximation) the Bayes Factors that are required to implement the framework in practice. However, this lies outside the scope of this paper, and instead we simply illustrate via simulations and
real data analysis the potential power gains of multivariate analysis.

Figure \ref{sfig:4vs2}a and \ref{sfig:4vs2}b shows results of simulations, based on both HDMP and NFBC1966 data, comparing power of the 
multivariate LRT of all four phenotypes vs conducting all six two-phenotype analyses and applying a Bonferroni correction for the six tests performed.
In these simulations the four-phenotype analysis is consistently more powerful (or as powerful) as the two-phenotype analyses, even when only one or two of the four phenotypes are truly associated with genotype. (Without Bonforroni correction the two-phenotype analysis is slightly more powerful for two out of eight simulations, and less powerful for the other six; Figure \ref{sfig:4vs2}c and \ref{sfig:4vs2}d.) While it may seem counter-intuitive that a four-phenotype analysis is more powerful than a two-phenotype analysis even when exactly two phenotypes are associated with genotype, this is actually expected, for reasons discussed in \cite{Stephens:2013}: including unassociated phenotypes in the multivariate analysis can increase power if these unassociated phenotypes are correlated with the associated phenotypes. 

We also applied four-phenotype, two-phenotype, and univariate analyses to the NFBC1966 data. 
In total, 45 SNPs from 14 genetic regions pass a significance level of 0.05 after Bonferroni correction (both for the number of SNPs and, in univariate and two-phenotype analyses, for the number of tests) in either the four-phenotype, two-phenotype, or univariate analyses. 
As expected, some SNPs show stronger signals in the four-phenotype analysis, whereas others show stronger signals in a two-phenotype or univariate analysis. Comparing the
four-phenotype analysis with the univariate analysis (Table \ref{tab:p_nfbc_s}), 16 SNPs
were significant in the four-phenotype analysis and not the univariate analysis;
whereas 3 SNPs were significant only in the univariate analysis.  
Comparing the four-phenotype analysis with the two-phenotype analysis (Table \ref{tab:p_nfbc_m}), 1 SNP was significant in the four-phenotype analysis and not the two-phenotype analysis, whereas no SNP was significant only in the two-phenotype analysis. 

These simulation and real-data results are consistent with the idea that multivariate tests will often provide more power than multiple univariate or pairwise tests. However, it is also clear that in a GWAS setting no single test will be the most powerful to detect the many different types
of genetic effects that could occur. Indeed, as in \cite{Stephens:2013}, it should be possible to manufacture simulations so that any
given test is the most powerful. Therefore we prefer
to emphasize that different multivariate and univariate tests can be complementary to one another, rather than competing.

\subsubsection*{Missing phenotype imputation}

One limitation of our algorithms is that they require
fully observed phenotypes. Since in a typical study many individuals may have partially missing phenotypes, removing all such individuals could substantially reduce power. To address this we developed
a phenotype imputation scheme for mvLMMs (Supplementary Note), 
which can be applied
to impute missing phenotypes before applying our LRT methods.
In brief, the imputation method first estimates parameters of the mvLMM under the null model using individuals with fully observed phenotypes, and then, conditional on these estimates and the observed phenotype data, imputes missing phenotypes using their conditional means.
Figure \ref{sfig:impute} shows the results of simulations, based on both HMDP and NFBC1966 data, comparing the power of this imputation-based approach with the alternative approach of dropping individuals with
partially missing phenotypes. For the HMDP simulations, because of the high relatedness, both methods achieve almost identical power as if all phenotypes are fully observed. For the NFBC1966 data, phenotype imputation achieves consistently greater power than dropping individuals, and in many simulation scenarios achieves power similar to that achieved if all phenotypes are observed (0\% missingness in Figure \ref{sfig:impute}).

\section*{Discussion}
Here, we present novel algorithms, and a software implementation in the package GEMMA, for multivariate analysis using mvLMMs in genetic association studies. This method is the first computationally-practical method for computing the LRT for the mvLMMs in GWAS applications, and the first
software implementation of any kind of test for GWAS with more than two phenotypes. Through simulation, we demonstrated that the $p$ values from  
our test are well-calibrated under the null, unlike $p$ values from existing
methods based on an approximation to the LRT.

Our algorithms are not without their limitations. Perhaps the most
fundamental is that, 
like its univariate counter-parts, our algorithms only apply to mvLMMs with one variance component (in addition to the residual error term). However, additional assumptions may allow our algorithms to be
extended to allow for more variance components \cite{Kostem:2013}. 
In addition, although our implementation of the EM algorithm scales only quadratically with the number of phenotypes, $d$, and so could be applied to reasonably large $d$, in practice
we caution that there could remain both computational and statistical barriers to applying these methods to even quite modest values of $d$ (e.g.~$d \approx 10$). Computationally, 
the number of iterations required to converge for larger $d$ will
inevitably increase, and ultimately this issue could be
the main barrier to effectively maximizing the likelihood for large $d$. 
Statistically, the number of parameters in the mvLMM is also quadratic in the number of phenotypes (the number of parameters in the two variance components is $d(d+1)$). Therefore, with a moderate sample size, additional assumptions on the structure of the variance components may be necessary to obtain reliable estimates. For example, one could penalize the departure of the estimated variance components to some prior values \cite{Meyer:2010}, or could assume the variance components are of low rank \cite{Kirkpatrick:2004} or sparse \cite{Runcie:2013}. The effectiveness of such strategies, paired with the mvLMM, in GWASs is a topic for potential
further research.

The most computationally expensive part of our method, as in the univariate case, is the eigen-decomposition step. (Although the genotype transformation step can be expensive with a large number of markers, it can be easily parallelized in a computing cluster.) The initial eigen-decomposition step not only requires a large amount of physical memory, but also becomes computationally intractable in practice, for large $n$ (e.g. $>50,000$ \cite{Zheng:2012}). Low rank approximations to the relatedness matrix \cite{Lippert:2011, Listgarten:2012, Zhou:2012} can alleviate both computation and memory requirements, and using these kinds of methods mvLMMs could be applied to very large GWASs.

\clearpage
\newpage

\section*{Methods}
\subsection*{Genotype and Phenotype Data}
We analyzed two data sets: the Hybrid Mouse Diversity Panel (HMDP) \cite{Bennett:2010} and the Northern Finland Birth Cohort 1966 (NFBC1966) Study \cite{Sabatti:2008}. 

The HMDP data includes 100 inbred strains with four phenotypes (high-density lipoprotein, HDL; total cholesterol, TC; triglycerides, TG; unesterified cholesterol, UC) and four million high quality fully imputed SNPs (SNPs are downloaded from \url{http://mouse.cs.ucla.edu/mousehapmap/full.html}). We excluded mice with missing phenotypes for any of these four phenotypes. We excluded non-polymorphic SNPs, and SNPs with a minor allele frequency less than 5\%. For SNPs that have identical genotypes, we tried to retain only one of them (by using ``--indep-pairwise 100 5 0.999999" option in PLINK \cite{Purcell:2007}). This left us with 98 strains, 656 individuals and 108,562 SNPs. We quantile transformed each phenotype to a standard normal distribution to guard against model mis-specification. We used the product of centered genotype matrix as an estimate of relatedness \cite{Hayes:2009, Astle:2009, Zhou:2013}. Note that the sample size used here is smaller than the original study \cite{Bennett:2010}, and the phenotypes are quantile-transformed instead of log transformed for robustness.

The NFBC1966 data contains 5402 individuals with multiple metabolic traits measured and 364,590 SNPs typed. We selected four phenotypes (high-density lipoprotein, HDL; low-density lipoprotein, LDL; triglycerides, TG; C-reactive protein, CRP) among them, following previous studies \cite{Korte:2012}. We selected individuals and SNPs following previous studies \cite{Sabatti:2008, Kang:2010} with the software PLINK \cite{Purcell:2007}. Specifically, we excluded individuals with missing phenotypes for any of these four phenotypes or having discrepancies between reported sex and sex determined from the X chromosome. We excluded SNPs with a minor allele frequency less than 1\%, having missing values in more than 1\% of the individuals, or with a Hardy-Weinberg equilibrium $p$ value below 0.0001. This left us with 5,255 individuals and 319,111 SNPs. For each phenotype, we quantile transformed the phenotypic values to a standard normal distribution, regressed out sex, oral contraceptives and pregnancy status effects \cite{Sabatti:2008}, and quantile transformed the residuals to a standard normal distribution again. We replaced the missing genotypes for a given SNP with its mean genotype value. We used the product of centered and scaled genotype matrix as an estimate of relatedness \cite{Hayes:2009, Astle:2009, Kang:2010}. 

In both data sets, we quantile transformed each single phenotype to a standard normal distribution to guard against model misspecification. Although this strategy does not guarantee that the transformed phenotypes follow a multivariate normal distribution jointly, it often works well in practice when the number of phenotypes is small (see, e.g. \cite{Stephens:2013}). For both data sets, we used a standard mvLMM with an intercept term (without any other covariates), and test each SNP in turn. Because the software MTMM relies on the commercial software ASREML to estimate the variance components in the null model, we modified the MTMM source code so that it can read in the estimated variance components from GEMMA.

\subsection*{Simulations}
To check if GEMMA and MTMM produce calibrated $p$ values, we used real genotypes in the HMDP data and simulated phenotypes under the null. Specifically, we simulated 10,000 null phenotypes based on the real relatedness matrix and the two estimated variance components (for HDL and TG), and calculated $p$ values for each SNP-phenotype pair (approximately 1 trillion pairs) in turn. We did not perform comparisons based on the NFBC1966 data, partly because GEMMA and MTMM produce identical $p$ values there, and partly because the sample size in NFBC1966 makes it computationally impractical to perform trillions of association tests to check for the type I error at the genome-wide significance level. 

To compare power between GEMMA and MTMM, we used real genotypes from the HMDP and NFBC1966 data, and we simulated phenotypes by adding genotype effects back to the original phenotypes \cite{Zhang:2010, Zhou:2012}. Specifically, we first identified SNPs unassociated with the four phenotypes based on one-phenotype, two-phenotype and four-phenotype analyses (LRT $p$ value $> 0.1$ in any of the 11 tests). We ordered the SNPs (76,780 in HMDP and 208,145 in NFBC1966) satisfying this criteria by their genomic location, and selected from them 10,000 evenly spaced SNPs to act as causal SNPs. For each causal SNP, we specified its effect size for the first trait (HDL) to explain a particular percentage of the phenotypic variance (proportion of variance explained, or PVE). Afterwards, we specified its effect for the second trait (TG) so that the proportion of variance in the second trait explained by the SNP equals to either 20\% or 80\% of the PVE in the first trait. We considered effect sizes for the two traits to be either in the same direction or in the opposite directions, and we added the simulated effects back to the original phenotypes to form the new simulated phenotypes. For each pre-specified PVE (ranged from 2\% to 20\% in HMDP and 0.04\% to 0.4\% in NFBC1966), we simulated 10000 sets of phenotypes, one for each causal SNP, and calculated $p$ values for each SNP-phenotype pair. We calculated statistical power as the proportion of $p$ values exceeding the genome-wide significance level at the conventional 0.05 level after Bonferroni correction ($p=4.6\times 10^{-7}$ for HDMP and $p=1.6\times 10^{-7}$ for NFBC1966). Notice that we simulated phenotypes based on HDL and TG in both data sets, and the two phenotypes are positively correlated in HMDP but negatively correlated in NFBC1966.

Our algorithms rely on fully observed phenotypes, and when missing phenotypes are present, we have to either drop individuals with partially missing phenotypes, or impute these missing values. To compare power between the two approaches (imputation versus dropping), we used the same set of simulated phenotypes described above, but randomly made 2.5\%, 5\% or 10\% of the individuals to have one phenotype missing. We calculated $p$ values for each SNP-phenotype pair from the two approaches using GEMMA, and calculated statistical power at the conventional 0.05 level after Bonferroni correction. 

Finally, we performed a power comparison between the four-phenotype analysis and the two-phenotype analysis using GEMMA, with simulations based on the two data sets. Specifically, we used the same set of 10,000 SNPs described above to act as causal SNPs, and we simulated phenotypes by adding genotype effects to the observed phenotypes, as above. For each causal SNPs, we made it to affect either one, two, three or four phenotypes. When the causal SNP affected two or four phenotypes, its effects on randomly selected half of the traits were in the opposite direction as that on the other half. When the causal SNP affected three phenotypes, its effects on randomly selected two traits were in the opposite direction as that on the third trait. The SNP effect size for each affected phenotype was simulated independently to account for a pre-specified PVE of that phenotype (ranged from 0.5\% to 5\% in HMDP and 0.04\% to 0.4\% in NFBC1966), which was further scaled with a random factor draw from a uniform distribution $\mbox{U}(0.8, 1)$. The simulated effects were added back to the original phenotypes to form the new simulated phenotypes. For the four-phenotype analysis, we calculated $p$ values for each SNP-phenotype pair and we calculated statistical power at the conventional 0.05 level after Bonferroni correction ($p=4.6\times 10^{-7}$ for HDMP and $p=1.6\times 10^{-7}$ for NFBC1966). For the two-phenotype analysis, we obtained the minimal $p$ value from the six pair-wise analyses for each SNP, and calculated statistical power as the proportion of these $p$ values exceeding either the same significance level ($p=4.6\times 10^{-7}$ for HDMP and $p=1.6\times 10^{-7}$ for NFBC1966), or a significance level that was further adjusted to account for the six tests performed ($p=7.6\times 10^{-8}$ for HDMP and $p=2.6\times 10^{-8}$ for NFBC1966).

\section*{Acknowledgment}
This research is supported in part by NIH grant HL092206 (PI Y Gilad) and NIH grant HG02585 to MS. We thank A.~J.~Lusis for making the mouse genotype and phenotype data available. We thank the NFBC1966 Study Investigators for making the NFBC1966 data available. The NFBC1966 Study is conducted and supported by the National Heart, Lung, and Blood Institute (NHLBI) in collaboration with the Broad Institute, UCLA, University of Oulu, and the National Institute for Health and Welfare in Finland. This manuscript was not prepared in collaboration with investigators of the NFBC1966 Study and does not necessarily reflect the opinions or views of the NFBC1966 Study Investigators, Broad Institute, UCLA, University of Oulu, National Institute for Health and Welfare in Finland and the NHLBI.

\clearpage
\newpage

\section{Supplementary Note}

\subsection{Multivariate Linear Mixed Model}
We consider the multivariate linear mixed model \cite{Henderson:1984},
\begin{equation}
\tilde{\bY}=\bA\tilde{\bW}+\bbeta\tilde{\bx}^T+\tilde{\bG}+\tilde{\bE}; \quad \tilde{\bG} \sim \mbox{MN}_{d \times n}({\bf 0}, \bV_g, \bK), \quad \tilde{\bE} \sim \mbox{MN}_{d \times n}({\bf 0}, \bV_e, \bI_{n\times n}),
\end{equation}
where $n$ is the number of individuals, $d$ is the number of phenotypes, $\tilde{\bY}$ is a $d$ by $n$ matrix of phenotypes, $\tilde{\bW}$ is a $c$ by $n$ matrix of covariates including a row of $1$s as intercept and $\bA$ is a $d$ by $c$ matrix of corresponding coefficients, $\tilde{\bx}$ is a $n$-vector of genotype for a particular marker and $\bbeta$ is a $d$-vector of its effect sizes for the $d$ phenotypes, $\tilde{\bG}$ is a $d$ by $n$ matrix of random effects, $\tilde{\bE}$ is a $d$ by $n$ matrix of residual errors, $\bK$ is a known $n$ by $n$ relatedness matrix, $\bI_{n\times n}$ is a $n$ by $n$ identity matrix, $\bV_g$ is a $d$ by $d$  symmetric matrix of genetic variance component, $\bV_e$ is a $d$ by $d$  symmetric matrix of environmental variance component and $\mbox{MN}_{d \times n}(\bf{0}, \bV_1, \bV_2)$ denotes the $d \times n$ matrix normal distribution with mean 0, row covariance matrix $\bV_1$ ($d$ by $d$), and column covariance matrix $\bV_2$ ($n$ by $n$). 

We group all covariates together into a $(c+1)$ by $n$ matrix $\tilde{\bX}=\begin{pmatrix} \tilde{\bW} \\ \tilde{\bx}^T \end{pmatrix}$, and group all coefficients together into a $d$ by $(c+1)$ matrix $\bB=(\bA, \bbeta)$.

Following \cite{Lippert:2011, Zhou:2012, Pirinen:2013}, we perform an eigen-decomposition of the relatedness matrix $\bK=\bU_k\bD_k\bU_k^T$, where $\bU_k$ is a $n$ by $n$ matrix of eigen vectors and $\bD_k$ is a diagonal $n$ by $n$ matrix filled with the corresponding eigen values, or diag($\delta_1, \cdots, \delta_n$). We then obtain transformed phenotype matrix $\bY=\tilde{\bY}\bU_k$ and transformed covariate matrix $\bX=\tilde{\bX}\bU_k$. We further denote $\bG=\tilde{\bG}\bU_k$ as the transformed random effect matrix, and $\bE=\tilde{\bE}\bU_k$ as the transformed residual error matrix. Now, the transformed phenotypes given the transformed covariates follow
\begin{equation}
\bY=\bB\bX+\bG+\bE; \quad \bG\sim \mbox{MN}(0, \bV_g, \bD_k), \quad \bE\sim \mbox{MN}(0, \bV_e, \bI_{n\times n}),
\end{equation}
which is equivalent to
\begin{equation}
\by=\bX^T\otimes \bI_{d\times d} \bb+\bg+\be; \quad \bg\sim \mbox{MVN}(0, \bD_k\otimes \bV_g), \quad \be\sim \mbox{MVN}(0, \bI_{n\times n}\otimes \bV_e),
\end{equation}
where $\by=\mbox{vec}(\bY)$, $\bb=\mbox{vec}(\bB)$, $\bg=\mbox{vec}(\bG)$, $\be=\mbox{vec}(\bE)$, vec denotes vectorization (i.e. stacking columns), MVN denotes multivariate normal distribution and $\otimes$ denotes Kronecker product.

Therefore, for each individual $l$, the transformed phenotypes given the transformed covariates follow independent (but not identical) multivariate normal distributions
\begin{equation}
\by_l=\bB\bx_l+\bg_l+\be_l; \quad \bg_l \sim \mbox{MVN}(0, \delta_l \bV_g), \quad \be_l \sim \mbox{MVN}(0, \bV_e),
\end{equation}
with variance $\bV_l=\delta_l\bV_g+\bV_e$, where $\by_l$ is the $l$th column vector of $\bY$, $\bx_l$ is $l$th column vector of $\bX$, $\bg_l$ is $l$th column vector of $\bG$, $\be_l$ is $l$th column vector of $\bE$, $\forall l=1,\cdots, n$.


\subsection{Optimization Method Overview}
We are interested in obtaining parameter estimates from this model, which are used further to obtain statistics and $p$ values to test the null hypothesis that the marker effect sizes for all phenotypes are zero, $H_0: \bbeta= \bf 0$, where $\bf 0$ is a $d$-vector of zeros, against the general alternative $H_1: \bbeta\neq \bf 0$. This test, in the bivariate case, corresponds to the ``full test" in MTMM. We do not consider either the ``interaction test" or the ``common test" in MTMM here. 

Parameter estimation in a mvLMM presents substantial computational challenges, in part because it requires multi-dimensional optimization for a potentially non-convex function. Procedures for multi-dimensional optimization can be classified into two categories based on whether or not they use derivatives. Derivative-free methods evaluate the (restricted) likelihood function for every combination of parameters along a searching path \cite{Graser:1987, Meyer:1989, Meyer:1991}. They are easy to implement, but are often computationally inefficient: their time complexity grows exponentially with the number of parameters, making them impractical for a reasonably large number of phenotypes \cite{Meyer:1996}. (For instance, the original paper on the derivative-free method for mvLMM only showed examples for two phenotypes \cite{Meyer:1991}.) The derivative-based methods include the expectation maximization (EM) algorithm \cite{Dempster:1977} and its accelerated version using parameter expansion (PX-EM) \cite{Liu:1998, Foulley:2000}; and the Newton-Raphson (NR) algorithm \cite{Patterson:1971, Thompson:1973} and its variant, the average information (AI) algorithm \cite{Gilmour:1995}. Because of the stability of EM-type algorithms (each iteration is guaranteed to increase the likelihood), and the faster convergence rate of NR-type algorithms, the two are often combined to gain the benefits of both (e.g.~PX-AI algorithm) \cite{Meyer:2006}. This strategy is used in many existing software packages, including the free packages GCTA \cite{Yang:2011b, Lee:2012b}, and WOMBAT \cite{Meyer:2007}, and the commercial package ASREML \cite{Gilmour:1995}. 

Unfortunately, even with the PX-AI algorithm, the per-iteration computation time for fitting a mvLMM still increases cubically, or worse, both with the number of individuals ($n$) and with the number of phenotypes ($d$) (the computational complexity is $O(n^3d^3)$ for EM and $O(n^3d^7)$ for AI). This is because existing method require repeated ``inversion" (actually, solving a system of linear equations) of an $nd \times nd$ matrix, in every iteration of the EM-like algorithm, and for evaluating every element inside the average information matrix (which is a $d(d+1)$ by $d(d+1)$ matrix) during each iteration of the NR-like algorithm, a computationally expensive procedure which increases cubically with both $n$ and $d$ ($O(n^3 d^3)$). This becomes especially problematic in GWASs where the optimizations are performed for every SNP in turn. To address this issue, \cite{Korte:2012} recently introduced the multi-trait mixed model (MTMM) method \cite{Korte:2012}, implemented in the MTMM software, to use an approximation strategy \cite{Aulchenko:2007, Zhang:2010, Kang:2010} to reduce computation time from cubic to quadratic in $n$. Specifically, the approximation avoids repeatedly re-optimizing the variance components under the alternative model for each SNP, by re-using part of the pre-estimated variance components under the null model (fit using the software ASREML) to approximate the likelihood ratio statistic. However, the approximated likelihood ratio statistic $2l_1(\tilde \bbeta_1, \tilde \btheta_1', \hat \btheta_0)-2l_0(\hat \btheta_0', \hat \btheta_0)$ is guaranteed to be less than the true likelihood ratio statistic $2l_1(\hat\bbeta_1, \hat \btheta_1', \hat \btheta_1)-2l_0(\hat \btheta_0', \hat \btheta_0)$, since $l_1(\tilde \bbeta_1, \tilde \btheta_1', \hat \btheta_0)<l_1(\hat\bbeta_1, \hat \btheta_1', \hat \btheta_1)$, where $\btheta$ are the variance component parameters to be fixed from $H_0$, $\btheta'$ are the variance component parameters to be estimated from $H_1$, $\hat \btheta$, $\hat \btheta'$ and $\hat \bbeta$ are MLE estimates, while $\tilde \btheta_1'$ and $\tilde \bbeta$ are conditional MLE estimates given $\hat \btheta_0$.

Here, we present novel algorithms that substantially reduce the computation burden. Our algorithms combine recently described univariate LMM tricks \cite{Lippert:2011, Zhou:2012, Pirinen:2013}, with the simultaneous diagonalization (known as the canonical transformation in animal breeding literatures \cite{Meyer:1991, Ducrocq:1997}) for the PX-EM algorithm, and with a few block-diagonal matrix and sparse matrix properties for the NR algorithm. In effect, our algorithms provide the multivariate analogue of the univariate algorithms EMMA \cite{Kang:2008}, and FaSTLMM/GEMMA/CM \cite{Lippert:2011, Zhou:2012, Pirinen:2013}. Specifically, with one $O(n^3)$ operation upfront, 
\begin{enumerate}
\item The EMMA algorithm reduced the computational cost per iteration for a {\it single} univariate LMM ($d=1$) from $O(n^3)$ to $O(n)$; in the multivariate case our algorithms reduce $O(n^3 d^3)$ to $O(nd^2)$ for EM and reduce $O(n^3 d^7)$ to $O(nd^6)$ for NR . 
\item FaSTLMM/GEMMA/CM reduced the computation cost per SNP for univariate LMMs from $O(n^3)$ to $O(n^2)$ (or $O(n)$ if $\bK$ has low rank\cite{Lippert:2011, Listgarten:2012}); in the multivariate case our algorithms reduce $O(n^3 d^3)$ per SNP to $O(n^2)$ (or $O(n)$ if $\bK$ has low rank).
\end{enumerate}

Our algorithms also obviate the need for the widely used AI algorithm \cite{Gilmour:1995} because our implementation of the NR algorithm has the same time complexity and practical computation time.  

For numerical optimization in the null model, we initialize the two variance components to be both diagonal matrices, with diagonal elements estimated from the corresponding univariate LMMs. We then perform the PX-EM algorithm, as described in details below, for 10,000 iterations or until the log likelihood increase between two consecutive iterations is below $10^{-4}$. Afterwards, we perform the NR algorithm, as described in details below,  using variance component estimates from the previous PX-EM algorithm, for another 100 iterations or until the log likelihood increase between two consecutive iterations is below $10^{-4}$. For GWAS applications, for each SNP tested, we use the variance components estimated from the null model as initial values. Because for moderate $d$ the PX-EM algorithm is considerably faster than the NR algorithm, we perform the NR algorithm only for markers where the $p$ value after the PX-EM algorithm is below $1.0\times 10^{-3}$. With the above thresholds, it often takes hundreds to thousands PX-EM iterations followed by a dozen NR iterations to optimize the null model, and often takes a few dozen PX-EM iterations followed by a couple NR iterations to optimize the alternative model for each SNP. Notice that all the precision thresholds and maximal iterations listed above can be adjusted in GEMMA. 

In our experience, both the PX-EM and NR algorithms are required for optimization, consistent with previous observations (e.g. \cite{Gilmour:1995, Meyer:2006, Meyer:2007, Yang:2011b, Lee:2012b}). On one hand, although the PX-EM algorithm is stable (each iteration is guaranteed to increase the likelihood), it is often too slow to converge. For example, in many data sets we have worked on, even thousands of PX-EM iterations may not be sufficient to maximize the null model (e.g. the difference between the resulting log likelihood and the true maximal value can be as large as one). And in many GWAS applications, using the PX-EM algorithm alone will fail to optimize the alternative model in a large fraction of SNPs (an average of 83\% of the $p$ values from the pairwise analyses using PX-EM alone are closer to the MTMM $p$ values than to the $p$ values obtained by the PX-EM plus NR algorithms in the HMDP data set; Figure \ref{sfig:pxemvsboth}). On the other hand, the NR algorithm is fast to converge given a good initial value (only takes a few iterations), but can easily fail to do so given a bad starting point. Therefore, we follow previous approaches \cite{Gilmour:1995, Meyer:2006, Meyer:2007, Yang:2011b, Lee:2012b} and combine the two algorithms together, with the PX-EM algorithm providing a good starting value for the following NR algorithm. In addition, for moderate $d$ the PX-EM algorithm is considerably faster than the NR algorithm, and so for GWAS applications, we perform the NR algorithm only for markers where the $p$ value after the PX-EM algorithm is $<1.0\times 10^{-3}$ (or a user adjusted threshold). This strategy makes GWAS analysis a few times faster than using NR algorithm for every marker, without noticeable loss of accuracy.

\subsection{PX-EM Algorithms}
Here, we describe an expectation conditional maximization (ECM) algorithm \cite{Meng:1993} for finding maximum likelihood estimates (MLE) in mvLMM, an expectation maximization (EM) algorithm \cite{Dempster:1977} for finding restricted maximum likelihood estimates (REMLE) in mvLMM, parameter expansion (PX) versions \cite{Liu:1998, Foulley:2000} of the two, and their efficient computations. 

\subsubsection{An ECM Algorithm for MLE}
We view $\bG$ as missing values, and we have the joint likelihood function as
\begin{equation}
\label{em_logl}
\log l(\bY, \bG | \bB, \bV_g, \bV_e)=\sum_{l=1}^n \{-d\log(2\pi)-\frac{1}{2}\log|\bV_e|-\frac{1}{2}\log|\delta_l\bV_g|-\frac{1}{2}\be_l^T\bV_e^{-1}\be_l-\frac{1}{2}\bg_l^T(\delta_l\bV_g)^{-1}\bg_l \}.
\end{equation}
The conditional distribution of $\bG$ given $\bY$ and the current values of $\bB^{(t)}$, $\bV_g^{(t)}$, $\bV_e^{(t)}$ follows
\begin{equation}
\bg_l|\by_l, \bB^{(t)}, \bV_g^{(t)}, \bV_e^{(t)} \sim \mbox{MVN}(\hat \bg_l^{(t)}, \hat\bSigma_l^{(t)}),
\end{equation}
where $\bV_l^{(t)}=\delta_l\bV_g^{(t)}+\bV_e^{(t)}$, $\hat\bg_l^{(t)}=\delta_l\bV_g^{(t)}(\bV_l^{(t)})^{-1}(\by_l-\bB^{(t)}\bx_l)$ and $\hat\bSigma_l^{(t)}=\delta_l\bV_g^{(t)}(\bV_l^{(t)})^{-1}\bV_e^{(t)}$. 

The expected value of the log likelihood function, with respect to the conditional distribution of $\bG$ given $\bY$ and the current values of $\bB^{(t)}$, $\bV_g^{(t)}$, $\bV_e^{(t)}$, is 
\begin{align}
&E_{\bG|\bY, \bB^{(t)}, \bV_g^{(t)}, \bV_e^{(t)}}[\log l(\bY, \bG | \bB, \bV_g, \bV_e)]\nonumber \\
&=\sum_{l=1}^n\{ -d\log(2\pi)-\frac{1}{2}\log|\delta_l\bV_g|-\frac{1}{2}\log|\bV_e|-\frac{1}{2}(\by_l-\bB\bx_l)^T\bV_e^{-1}(\by_l-\bB\bx_l) \nonumber \\
&-\frac{1}{2}(\hat\bg_l^{(t)})^T((\delta_l\bV_g)^{-1}+\bV_e^{-1})^{-1}\hat\bg_l^{(t)}-\frac{1}{2}\mbox{trace}( ( (\delta_l\bV_g)^{-1}+\bV_e^{-1})^{-1}\hat\bSigma_l^{(t)})+\hat\bg_l^{(t)}\bV_e^{-1}(\by_l-\bB\bx_l)  \}.
\end{align}
We optimize the above expectation using two conditional maximization steps, in which $\bB^{(t+1)}$ is updated conditional on $\bV_g^{(t)}$ and $\bV_e^{(t)}$, and $\bV_g^{(t+1)}$, $\bV_e^{(t+1)}$ are updated conditional on $\bB^{(t+1)}$, $\bV_g^{(t)}$ and $\bV_e^{(t)}$, or
\begin{align}
\bB^{(t+1)}&=(\bY-\hat\bG^{(t)})\bX(\bX\bX^T)^{-1}, \\
\bV_g^{(t+1)}&=\frac{1}{n}\sum_{l=1}^n \delta_l^{-1}(\hat\bg_l^{(t)}(\hat\bg_l^{(t)})^T+\hat\bSigma_l^{(t)}), \\
\bV_e^{(t+1)}&=\frac{1}{n}\sum_{l=1}^n (\hat\be_l^{(t)}(\hat\be_l^{(t)})^T+\hat\bSigma_l^{(t)}),
\end{align}
where $\hat\bG^{(t)}$ is a $d$ by $n$ matrix with $l$th column $\hat\bg_l^{(t)}$, $\hat\be_l^{(t)}=\by_l-\bB^{(t+1)}\bx_l-\hat\bg_l^{(t)}$. We note that the derivation of the last two equations requires obtaining the partial derivatives with respect to $\mbox{vec}(\bV_g)$ and $\mbox{vec}(\bV_e)$ based on a few matrix calculus properties listed in \cite{Zhou:2012}.

\subsubsection{An EM Algorithm for REMLE}
We view both $\bB$ and $\bG$ as missing values. The joint likelihood function remains the same as in equation \ref{em_logl}, and the joint conditional distribution of $\bB$, $\bG$ given $\bY$ and the current values of $\bV_g^{(t)}$, $\bV_e^{(t)}$ is
\begin{align}
\left( {\begin{array} {c}
\bb \\
\bg \\
\end{array}} \right)  | \bY, \bV_g^{(t)}, \bV_e^{(t)} &\sim \mbox{MVN} (\left( {\begin{array} {cc}
\hat\bSigma_{\bb\bb}^{(t)} & \hat\bSigma_{\bb\bg}^{(t)}\\
\hat\bSigma_{\bg\bb}^{(t)} & \hat\bSigma_{\bg\bg}^{(t)}\\
\end{array}} \right)\left( {\begin{array} {c}
(\bX\otimes (\bV_e^{(t)})^{-1}) \by \\
(\bI_{n\times n} \otimes (\bV_e^{(t)})^{-1}) \by \\
\end{array}} \right) ,\left( {\begin{array} {cc}
\hat\bSigma_{\bb\bb}^{(t)} & \hat\bSigma_{\bb\bg}^{(t)}\\
\hat\bSigma_{\bg\bb}^{(t)} & \hat\bSigma_{\bg\bg}^{(t)}\\
\end{array}} \right) ), 
\end{align}
where
\begin{equation}
\left( {\begin{array} {cc}
\hat\bSigma_{\bb\bb}^{(t)} & \hat\bSigma_{\bb\bg}^{(t)}\\
\hat\bSigma_{\bg\bb}^{(t)} & \hat\bSigma_{\bg\bg}^{(t)}\\
\end{array}} \right)=\left( {\begin{array} {cc}
\bX\bX^T \otimes (\bV_e^{(t)})^{-1} & \bX \otimes (\bV_e^{(t)})^{-1}\\
\bX^T \otimes (\bV_e^{(t)})^{-1} & \bD_k^{-1}\otimes (\bV_g^{(t)})^{-1}+\bI_{n\times n}\otimes (\bV_e^{(t)})^{-1}\\
\end{array}} \right)^{-1}.
\end{equation}
Therefore, 
\begin{align}
\bg_l | \bY, \bV_g^{(t)}, \bV_e^{(t)}&\sim \mbox{MVN}(\hat\bg_l^{(t)}, \hat\bSigma_{l, \bg\bg}^{(t)} ),\\
\be_l=\by_l-\bB\bx_l-\bg_l | \bY, \bV_g^{(t)}, \bV_e^{(t)} & \sim \mbox{MVN}(\hat\be_l^{(t)}, \hat\bSigma_{l, \be\be}^{(t)} ),
\end{align}
where 
\begin{align}
\hat\bb^{(t)}&=\hat\bSigma_{\bb\bb}^{(t)}\sum_{l=1}^n\bx_k\otimes(\bV_l^{(t)})^{-1}\by_l,\\
\hat\bg_l^{(t)}&=\sum_{l=1}^n \delta_l\bV_g^{(t)}(\bV_l^{(t)})^{-1}(\by_l-\hat\bB^{(t)}\bx_l), \\
\hat\be_l^{(t)}&=\by_l-\hat\bB^{(t)}\bx_l-\hat\bg_l^{(t)},\\ 
\hat\bSigma_{\bb\bb}^{(t)} &=\sum_{l=1}^n \bx_l\bx_l^T\otimes (\bV_l^{(t)})^{-1},\\
\hat\bSigma_{l, \bg\bg}^{(t)} &=\delta_l\bV_g^{(t)}(\bV_l^{(t)})^{-1} +(\bx_l^T\otimes \delta_l\bV_g^{(t)}(\bV_l^{(t)})^{-1})(\sum_{l=1}^n \bx_l\bx_l^T\otimes \delta_l\bV_g^{(t)}(\bV_l^{(t)})^{-1})^{-1}(\bx_l\otimes \delta_l\bV_g^{(t)}(\bV_l^{(t)})^{-1}),\\
\hat\bSigma_{l, \be\be}^{(t)} &=\delta_l\bV_g^{(t)}(\bV_l^{(t)})^{-1}+(\bx_l^T\otimes (\bV_l^{(t)})^{-1})(\sum_{l=1}^n \bx_l\bx_l^T\otimes \delta_l\bV_g^{(t)}(\bV_l^{(t)})^{-1})^{-1}(\bx_l\otimes (\bV_l^{(t)})^{-1}),
\end{align}
and $\hat\bB^{(t)}$ is the matrix satisfies $\mbox{vec}(\hat\bB^{(t)})=\hat\bb^{(t)}$. 

The expected value of the log likelihood function, with respect to the conditional distribution of $\bB$, $\bG$ given $\bY$ and the current values of $\bV_g^{(t)}$, $\bV_e^{(t)}$, is 
\begin{align}
&E_{\bB, \bG|\bY, \bV_g^{(t)}, \bV_e^{(t)}}[\log l(\bY, \bG, \bB| \bV_g, \bV_e)]\nonumber \\
&=\sum_{l=1}^n\{ -d\log(2\pi)-\frac{1}{2}\log|\bV_e|-\frac{1}{2}\log|\delta_l\bV_g|-\frac{1}{2}(\hat\be_l^{(t)})^T\bV_e^{-1}\hat\be_l^{(t)}-\frac{1}{2}\mbox{trace}( \bV_e^{-1}\hat\bSigma_{l,\be\be}^{(t)}) \nonumber \\
&-\frac{1}{2}(\hat\bg_l^{(t)})^T(\delta_l\bV_g)^{-1}\hat\bg_l^{(t)}-\frac{1}{2}\mbox{trace}( (\delta_l\bV_g)^{-1}\hat\bSigma_{l,\bg\bg}^{(t)})  \}.
\end{align}
We update $\bV_g^{(t+1)}$ and $\bV_e^{(t+1)}$ to maximize the above expectation
\begin{align}
\bV_g^{(t+1)}&=\frac{1}{n}\sum_{l=1}^n \delta_l^{-1}(\hat\bg_l^{(t)}(\hat\bg_l^{(t)})^T+\hat\bSigma_{l, \bg\bg}^{(t)}), \\
\bV_e^{(t+1)}&=\frac{1}{n}\sum_{l=1}^n (\hat\be_l^{(t)}(\hat\be_l^{(t)})^T+\hat\bSigma_{l, \be\be}^{(t)}).
\end{align}

\subsubsection{PX versions of ECM and EM}
We introduce a latent parameter $\bV_a$ as a $d$ by $d$ matrix to ensure $\bV_g=\bV_a\bV_g^*\bV_a$. The mvLMM with the new parameterization becomes
\begin{equation}
\by_l=\bB\bx_l+\bV_a\bg_l^*+\be_l \quad \bg_l^* \sim \mbox{MVN}(0, \delta_l \bV_g^*) \quad \be_l \sim \mbox{MVN}(0, \bV_e),
\end{equation}
The expectation of the log likelihood function in the ECM or the EM algorithm is taken at $\bV_a^{(t)}=\bI_{d\times d}$. The updates for $\bV_g^*$, $\bV_e$ (and $\bB$ for ECM) remain the same, and the update for $\bV_a$ is
\begin{align}
\bV_a^{(t+1)}&=(\frac{1}{n}\sum_{l=1}^n E((\by_l-\bB\bx_l)(\bg_l^*)^T))  (\frac{1}{n}\sum_{l=1}^n E(\bg_l^*(\bg_l^*)^T) )^{-1}.
\end{align}
where the expectations are taken with respect to the conditional distribution of $\bG^*$ (and $\bB$ for EM) given $\bY$ and the current values of $\bV_g^{(t)}$, $\bV_e^{(t)}$ (and $\bB^{(t+1)}$ for ECM).

\subsubsection{Efficient computation}
The most computationally expensive part of the ECM/EM algorithm is the evaluation of each $\bV_l^{-1}$ and further the calculation of quantities that involve these inverses. A naive brute force approach will make the computation cubic in the number of traits, which can be avoided, by performing a transformation that further converts correlated traits into uncorrelated ones (in addition to the transformation that we have already performed to convert correlated individuals into uncorrelated ones). The idea behind this is commonly referred to as the canonical transformation in animal breeding literatures (e.g. \cite{Meyer:1991, Ducrocq:1997} and references there in), or as the simultaneous diagonalization in linear algebra. 

More specifically, for each value of $\bV_g$ and $\bV_e$, we perform an eigen decomposition of the matrix $\bV_e^{-\frac{1}{2} }\bV_g \bV_e^{-\frac{1}{2} }=\bU_{\lambda}\bD_{\lambda}\bU_{\lambda}^T$, and we transform each phenotype vector $\by_l$ and each covariate vector $\bx_l$ by multiplying $\bU_{\lambda}^T \bV_e^{-\frac{1}{2} }$. After that, for each individual, the transformed phenotypes given the transformed genotypes will follow independent univariate normal distributions (rather than multivariate normal distributions). Subsequently, each $\bV_l^{-1}=\bV_e^{-\frac{1}{2} }\bU_{\lambda}(\bD_{\lambda}+\bI_{d\times d})^{-1}\bU_{\lambda}^T \bV_e^{-\frac{1}{2} }$ and quantities involving $\bV_l^{-1}$ can be calculated efficiently. 

Therefore, the computation complexity for each iteration in the (PX) ECM/EM algorithm is $O(nc^2d^2)$.

\subsection{Newton-Raphson's Algorithms}
Here, we describe Newton-Raphson's algorithms for MLE and REMLE estimation in mvLMM. Although an average-information algorithm \cite{Gilmour:1995} has often been used in place of a standard Newton-Raphson's algorithm, we found it unnecessary when we use the efficient algorithms described below.

\subsubsection{Target functions and partial derivatives}
Both the log-likelihood function and the log-restricted likelihood function can be expressed as functions for $\bV_g$ and $\bV_e$ only:
\begin{align}
l(\bV_g, \bV_e)&=-\frac{nd}{2}\log(2\pi)-\frac{1}{2} \log|\bH|-\frac{1}{2}\by^T\bP\by, \\
l_r(\bV_g, \bV_e)&=-\frac{(n-c-1)d}{2}\log(2\pi)+\frac{d}{2}\log|\bX\bX^T|-\frac{1}{2}\log|\bH| \nonumber \\
&-\frac{1}{2} \log|(\bX\otimes \bI_{d\times d}) \bH^{-1}(\bX^T\otimes \bI_{d\times d})  |-\frac{1}{2}\by^T\bP\by,
\end{align}
where 
\begin{align}
\bH&=\bD_k\otimes \bV_g+\bI_{n\times n}\otimes \bV_e=\mbox{diag}(\bV_l), \\
\bP&=\bH^{-1}-\bH^{-1}(\bX^T\otimes \bI_{d\times d}) ( (\bX\otimes \bI_{d\times d}) \bH^{-1}(\bX^T\otimes \bI_{d\times d}) )^{-1}(\bX\otimes \bI_{d\times d}) \bH^{-1}.
\end{align}
With a slight abuse of notation, we denote $v_{g,ij}$ as the $ij$th element of $\bV_g$, $v_{e,ij}$ as the $ij$th element of $\bV_e$, $\bI_{i}$ as a $d$-vector with $i$th element 1 and other elements 0, and $\bI_{ij}=\bI_i\bI_j^T$ as a $d$ by $d$ matrix with $ij$th element 1 and other elements 0. We have
\begin{align}
\frac{\partial \mbox{vec}(\bD_k\otimes \bV_g)}{\partial v_{g,ij}}&=\mbox{vec}(\bD_k\otimes (\bI_{ij}+\bI_{ji}) ) \frac{1}{1+1_{i=j}},\\
\frac{\partial \mbox{vec}(\bI_{n\times n}\otimes \bV_e)}{\partial v_{e,ij}}&=\mbox{vec}(\bI_{n\times n}\otimes (\bI_{ij}+\bI_{ji}) ) \frac{1}{1+1_{i=j}},
\end{align}
where $1_{i=j}$ is an indicator function that takes value 1 when $i$ equals $j$ and 0 otherwise. 

With a few matrix calculus properties listed in \cite{Zhou:2012}, we obtain the first order partial derivatives for the log-likelihood and the log-restricted likelihood functions
\begin{align}
\frac{\partial l }{\partial v_{g, ij} }&=\frac{1}{1+1_{i=j}}\{-\frac{1}{2}\mbox{trace}(\bH^{-1} (\bD_k\otimes (\bI_{ij}+\bI_{ji})) )+\frac{1}{2}\by^T\bP (\bD_k\otimes (\bI_{ij}+\bI_{ji})) \bP\by\}, \\
\frac{\partial l }{\partial v_{e, ij} }&=\frac{1}{1+1_{i=j}}\{-\frac{1}{2}\mbox{trace}(\bH^{-1} (\bI_{n\times n}\otimes (\bI_{ij}+\bI_{ji})) )+\frac{1}{2}\by^T\bP (\bI_{n\times n}\otimes (\bI_{ij}+\bI_{ji})) \bP\by\}, \\
\frac{\partial l_r }{\partial v_{g, ij} }&=\frac{1}{1+1_{i=j}}\{-\frac{1}{2}\mbox{trace}(\bP (\bD_k\otimes (\bI_{ij}+\bI_{ji}) ) )+\frac{1}{2}\by^T\bP (\bD_k\otimes (\bI_{ij}+\bI_{ji}) ) \bP\by \}, \\
\frac{\partial l_r }{\partial v_{e, ij} }&=\frac{1}{1+1_{i=j}}\{-\frac{1}{2}\mbox{trace}(\bP (\bI_{n\times n}\otimes (\bI_{ij}+\bI_{ji}) ) )+\frac{1}{2}\by^T\bP (\bI_{n\times n}\otimes (\bI_{ij}+\bI_{ji}) ) \bP\by \},
\end{align}
and the second order partial derivatives for the log-likelihood function
\begin{align}
\frac{\partial l^2 }{\partial v_{g, ij}\partial v_{g, i'j'}}&=\frac{1}{(1+1_{i=j})(1+1_{i'=j'})} \{ \frac{1}{2}\mbox{trace}(\bH^{-1} (\bD_k\otimes (\bI_{ij}+\bI_{ji}))\bH^{-1} (\bD_k\otimes (\bI_{i'j'}+\bI_{j'i'})) ) \nonumber\\
&-\by^T\bP (\bD_k\otimes (\bI_{ij}+\bI_{ji})) \bP (\bD_k\otimes (\bI_{i'j'}+\bI_{j'i'})) \bP\by \}, \\
\frac{\partial l^2 }{\partial v_{g, ij}\partial v_{e, i'j'}}&=\frac{1}{(1+1_{i=j})(1+1_{i'=j'})} \{ \frac{1}{2}\mbox{trace}(\bH^{-1} (\bD_k\otimes (\bI_{ij}+\bI_{ji}))\bH^{-1} (\bI_{n\times n}\otimes (\bI_{i'j'}+\bI_{j'i'})) ) \nonumber\\
&-\by^T\bP (\bD_k\otimes (\bI_{ij}+\bI_{ji})) \bP (\bI_{n\times n}\otimes (\bI_{i'j'}+\bI_{j'i'})) \bP\by \}, \\
\frac{\partial l^2 }{\partial v_{e, ij}\partial v_{e, i'j'}}&=\frac{1}{(1+1_{i=j})(1+1_{i'=j'})} \{ \frac{1}{2}\mbox{trace}(\bH^{-1} (\bI_{n\times n}\otimes (\bI_{ij}+\bI_{ji}))\bH^{-1} (\bI_{n\times n}\otimes (\bI_{i'j'}+\bI_{j'i'})) ) \nonumber\\
&-\by^T\bP (\bI_{n\times n}\otimes (\bI_{ij}+\bI_{ji})) \bP (\bI_{n\times n}\otimes (\bI_{i'j'}+\bI_{j'i'})) \bP\by \},
\end{align}
and second order partial derivatives for the log-restricted likelihood function
\begin{align}
\frac{\partial l_r^2 }{\partial v_{g, ij}\partial v_{g, i'j'}}&=\frac{1}{(1+1_{i=j})(1+1_{i'=j'})} \{ \frac{1}{2}\mbox{trace}(\bP (\bD_k\otimes (\bI_{ij}+\bI_{ji}))\bP (\bD_k\otimes (\bI_{i'j'}+\bI_{j'i'})) ) \nonumber\\
&-\by^T\bP (\bD_k\otimes (\bI_{ij}+\bI_{ji})) \bP (\bD_k\otimes (\bI_{i'j'}+\bI_{j'i'})) \bP\by \}, \\
\frac{\partial l_r^2 }{\partial v_{g, ij}\partial v_{e, i'j'}}&=\frac{1}{(1+1_{i=j})(1+1_{i'=j'})} \{ \frac{1}{2}\mbox{trace}(\bP (\bD_k\otimes (\bI_{ij}+\bI_{ji}))\bP (\bI_{n\times n}\otimes (\bI_{i'j'}+\bI_{j'i'})) ) \nonumber\\
&-\by^T\bP (\bD_k\otimes (\bI_{ij}+\bI_{ji})) \bP (\bI_{n\times n}\otimes (\bI_{i'j'}+\bI_{j'i'})) \bP\by \},\\
\frac{\partial l_r^2 }{\partial v_{e, ij}\partial v_{e, i'j'}}&=\frac{1}{(1+1_{i=j})(1+1_{i'=j'})} \{ \frac{1}{2}\mbox{trace}(\bP (\bI_{n\times n}\otimes (\bI_{ij}+\bI_{ji}))\bP (\bI_{n\times n}\otimes (\bI_{i'j'}+\bI_{j'i'})) ) \nonumber\\
&-\by^T\bP (\bI_{n\times n}\otimes (\bI_{ij}+\bI_{ji})) \bP (\bI_{n\times n}\otimes (\bI_{i'j'}+\bI_{j'i'})) \bP\by \}.
\end{align}

\subsubsection{Efficient computation}
We describe in this section the efficient calculations of the target functions, the first-order partial derivatives with respect to $v_{g, ij}$, and the second-order partial derivatives with respect to $v_{g, ij}$ and $v_{g, i'j'}$. The first-order and second-order partial derivatives with respect to other parameters can be calculated in a similar fashion. The calculations described here use basic properties of block diagonal matrices and sparse matrices.

We denote $\bQ=(\bX\otimes \bI_{d\times d}) \bH^{-1}(\bX^T\otimes \bI_{d\times d})$, $\bq=(\bX\otimes \bI_{d\times d}) \bH^{-1}\by$, $q=\by^T\bH^{-1}\by$, and with a slight abuse of notation, denote 
\begin{align}
\bQ_{ij}^{g}&=(\bX\otimes \bI_{d\times d}) \bH^{-1}(\bD_k\otimes \bI_{ij}) \bH^{-1} (\bX^T\otimes \bI_{d\times d}), \\
\bq_{ij}^{g}&=(\bX\otimes \bI_{d\times d}) \bH^{-1}(\bD_k\otimes \bI_{ij}) \bH^{-1} \by, \\
q_{ij}^{g}&=\by^T \bH^{-1}(\bD_k\otimes \bI_{ij}) \bH^{-1} \by, \\
\bQ_{ij, i'j'}^{gg}&=(\bX\otimes \bI_{d\times d}) \bH^{-1}(\bD_k\otimes \bI_{ij}) \bH^{-1}(\bD_k\otimes \bI_{i'j'}) \bH^{-1} (\bX^T\otimes \bI_{d\times d}), \\
\bq_{ij, i'j'}^{gg}&=(\bX\otimes \bI_{d\times d}) \bH^{-1}(\bD_k\otimes \bI_{ij})\bH^{-1}(\bD_k\otimes \bI_{i'j'}) \bH^{-1} \by, \\
q_{ij, i'j'}^{gg}&=\by^T \bH^{-1}(\bD_k\otimes \bI_{ij})\bH^{-1}(\bD_k\otimes \bI_{i'j'}) \bH^{-1} \by.
\end{align}
For the trace terms, we have
\begin{align}
\mbox{trace}&(\bP(\bD_k\otimes\bI_{ij}))=\mbox{trace}(\bH^{-1}(\bD_k\otimes\bI_{ij}))-\mbox{trace}(\bQ^{-1}\bQ_{ij}^{g} ), \\
\mbox{trace}&(\bP(\bD_k\otimes\bI_{ij})\bP(\bD_k\otimes\bI_{i'j'}))=\mbox{trace}(\bH^{-1}(\bD_k\otimes\bI_{ij})\bH^{-1}(\bD_k\otimes\bI_{i'j'}) ) \nonumber \\
&-\mbox{trace}(\bQ^{-1}\bQ_{ij, i'j'}^{gg} ) -\mbox{trace}(\bQ^{-1}\bQ_{i'j', ij}^{gg}  )+\mbox{trace}(\bQ^{-1}\bQ_{ij}^{g} \bQ^{-1}\bQ_{i'j'}^{g}).
\end{align}
For the vector-matrix-vector product terms, we have
\begin{align}
\by^T\bP\by=&q-\bq^T \bQ^{-1}\bq,\\
\by^T\bP(\bD_k&\otimes\bI_{ij})\bP\by=q_{ij}^{g}-\bq^T \bQ^{-1}\bq_{ij}^{g} -(\bq_{ij}^{g})^T \bQ^{-1}\bq+\bq^T\bQ^{-1}\bQ_{ij}^{g}\bQ^{-1}\bq, \\
\by^T\bP(\bD_k&\otimes\bI_{ij})\bP(\bD_k\otimes\bI_{i'j'})\bP\by=q_{ij, i'j'}^{gg}-\bq^T\bQ^{-1}\bq_{ij, i'j'}^{gg}-(\bq_{j'i', ji}^{gg})^T\bQ^{-1}\bq-(\bq_{ji}^{g})^T\bQ^{-1}\bq_{i'j'}^{g} \nonumber \\
& +\bq^T\bQ^{-1}\bQ_{ij}^{g}\bQ^{-1}\bq_{i'j'}^{g}+(\bq_{ji}^{g})^T\bQ^{-1}\bQ_{i'j'}^{g}\bQ^{-1}\bq+\bq^T\bQ^{-1}\bQ_{ij, i'j'}^{gg}\bq \nonumber\\
&-\bq^T\bQ^{-1}\bQ_{ij}^{g}\bQ^{-1}\bQ_{i'j'}^{g}\bQ^{-1}\bq.
\end{align}
Therefore, it suffices to efficiently evaluate
\begin{align}
|\bH|&=\sum_{l=1}^n |\bV_l|, \\
\mbox{trace}&(\bH^{-1}(\bD_k\otimes\bI_{ij}))=\sum_{l=1}^n \delta_l (\bI_j^T\bV_l^{-1}\bI_i), \\
\mbox{trace}&(\bH^{-1}(\bD_k\otimes\bI_{ij})\bH^{-1}(\bD_k\otimes\bI_{i'j'}))=\sum_{l=1}^n \delta_l^2 (\bI_{j'}^T\bV_l^{-1}\bI_i) (\bI_j^T\bV_l^{-1}\bI_{i'}),
\end{align}
and
\begin{align}
\bQ&=\sum_{l=1}^n (\bx_l\bx_l^T) \otimes \bV_l^{-1}, \\
\bq&=\sum_{l=1}^n \bx_l\otimes (\bV_l^{-1}\by_l), \\
q&=\sum_{l=1}^n \by_l^T\bV_l^{-1}\by_l,
\end{align}
and
\begin{align}
\bQ_{ij}^{g}&=\sum_{l=1}^n \delta_l (\bx_l\bx_l^T)\otimes (\bV_l^{-1}\bI_{ij}\bV_l^{-1}), \\
\bq_{ij}^{g}&=\sum_{l=1}^n \delta_l \bx_l \otimes (\bV_l^{-1}\bI_{ij}\bV_l^{-1} \by_l) , \\
q_{ij}^{g}&=\sum_{l=1}^n \delta_l \by_l^T \bV_l^{-1}\bI_{ij}\bV_l^{-1} \by_l, \\
\bQ_{ij, i'j'}^{gg}&=\sum_{l=1}^n \delta_l^2 (\bx_l\bx_l^T)\otimes (\bV_l^{-1}\bI_{ij}\bV_l^{-1}\bI_{i'j'}\bV_l^{-1}), \\
\bq_{ij, i'j'}^{gg}&=\sum_{l=1}^n \delta_l^2 \bx_l \otimes (\bV_l^{-1}\bI_{ij}\bV_l^{-1}\bI_{i'j'}\bV_l^{-1} \by_l), \\
q_{ij, i'j'}^{gg}&=\sum_{l=1}^n \delta_l^2 \by_l^T \bV_l^{-1}\bI_{ij}\bV_l^{-1}\bI_{i'j'}\bV_l^{-1} \by_l.
\end{align}
Notice that one key trick for calculating all the above quantities is observing that $\bI_{ij}=\bI_i\bI_j^T$. In this way, many of the above quantities only involve scalar multiplications or rank one matrix updates. 

The most time consuming part is the calculation of $\bQ_{ij, i'j'}^{gg}$, each requiring $O(nc^2d^2)$ computation time. The computation complexity for each iteration in the Newton-Raphson's algorithm is therefore $O(nc^2d^6)$.

\subsection{Test Statistics and $p$ Values}
\subsubsection{Test Statistics}
We consider three common tests for mvLMMs: the likelihood ratio test, the Wald test and the score test. 

The likelihood ratio test calculates the maximum log likelihoods for both the alternative ($\hat l_1(\hat\bV_{g, 1}, \hat\bV_{e, 1})$) and the null ($\hat l_0(\hat\bV_{g, 0}, \hat\bV_{e, 0})$) models. It computes a test statistics based on the difference between the two: $z_{LR}=2(\hat l_1(\hat\bV_{g,1}, \hat\bV_{e, 1})-\hat l_1(\hat\bV_{g, 0}, \hat\bV_{e, 0}))$. 

The Wald estimates the effect size
\begin{equation}
\hat \bbeta=\sum_{l=1}^n x_l\hat\bV_l^{-1}\by_l-(\sum_{l=1}^n (x_l\bw_l)\otimes \hat\bV_l^{-1} )(\sum_{l=1}^n (\bw_l\bw_l^T)\otimes \hat\bV_l^{-1})^{-1}(\sum_{l=1}^n \bw_l\otimes (\hat\bV_l^{-1}\by_l) ),
\end{equation}
and its precision $V(\hat \bbeta)^{-1}$
\begin{equation}
V(\hat \bbeta)^{-1}=\sum_{l=1}^n x_l^2\hat\bV_l^{-1}-(\sum_{l=1}^n (x_l\bw_l)\otimes \hat\bV_l^{-1} )(\sum_{l=1}^n (\bw_l\bw_l^T)\otimes \hat\bV_l^{-1})^{-1}(\sum_{l=1}^n (x_l\bw_l)\otimes \hat\bV_l^{-1} ),
\end{equation}
with the variance component estimates $\hat \bV_{g}$ and $\hat \bV_{e}$ ($\hat\bV_l=\delta_l\hat\bV_{g, 1}+\hat\bV_{e, 1}$) obtained under the alternative. Above, $x_l$ is the $l$th element of the transformed genotype vector $\bx$ and $\bw_l$ is the $l$th column vector of the transformed covariance matrix $\bW$. Afterwards, it computes $z_{Wald}=\sqrt{\hat\bbeta^TV(\hat \bbeta)^{-1}\hat\bbeta}$. 

The score test computes the score, a $d(d+1)$ vector $\hat\bs$, with each element equals to the corresponding first order partial derivate described in the previous sections, and the observed information matrix, a $d(d+1)$ by $d(d+1)$ matrix $\hat\bI$, whose formula is also described above. These values are evaluated with parameter estimates obtained under the null, and are used to compute $z_{Score}=\sqrt{\hat\bs^T\hat \bI^{-1}\hat\bs}$.

\subsubsection{$p$ Value Calibration}
Under the null hypothesis, all three test statistics follow a $\chi^2(d)$ distribution asymptotically. However, when the sample size is small {\it or} the relatedness structure is strong, then the test statistics will not follow the asymptotic distribution exactly, and $p$ values calculated from the asymptotic distribution will not be be calibrated (see, e.g. \cite{Evans:1982, Rothenberg:1984}). To correct for this, we follow \cite{Rothenberg:1984} and use Edgeworth-corrected critical values for the three tests. Specifically, the corrected $z$ scores from the three tests for a given marker is
\begin{align}
z_{LR}^c&=z_{LR}/(1+\frac{\hat a}{2d}), \\
z_{Wald}^c&=\frac{-(2d+\hat a+\hat b)(d+2)+\sqrt{(2d+\hat a+\hat b)^2(d+2)^2+8d(d+2)\hat c z_{Wald}} }{2\hat c}, \\
z_{Score}^c&=\frac{(2d+\hat a-\hat b)(d+2)-\sqrt{(2d+\hat a-\hat b)^2(d+2)^2-8d(d+2)\hat c z_{Score}} }{2\hat c},
\end{align}
where
\begin{align}
\hat a &= 2\hat b-\hat c,\\
\hat b &= 2\sum_{i<j}\sum_{i'<j'}\hat\Lambda_{ij, i'j'}\mbox{trace}((\bR\hat\bQ^{-1}\hat\bQ_{ij, i'j'}\hat\bQ^{-1}\bR^T-\bR\hat\bQ^{-1}\hat\bQ_{ij}\hat\bQ^{-1}\hat\bQ_{i'j'}\hat\bQ^{-1}\bR^T) (\bR\hat\bQ^{-1}\bR^T)^{-1}),\\
\nonumber \hat c &= \sum_{i<j}\sum_{i'<j'}\hat\Lambda_{ij, i'j'} (\mbox{trace}((\bR\hat\bQ^{-1}\hat\bQ_{ij}\hat\bQ^{-1}\bR^T) (\bR\hat\bQ^{-1}\bR^T)^{-1}(\bR\hat\bQ^{-1}\hat\bQ_{i'j'}\hat\bQ^{-1}\bR^T) (\bR\hat\bQ^{-1}\bR^T)^{-1}),\\ 
&+\frac{1}{2}\mbox{trace}((\bR\hat\bQ^{-1}\hat\bQ_{ij}\hat\bQ^{-1}\bR^T) (\bR\hat\bQ^{-1}\bR^T)^{-1})\mbox{trace}((\bR\hat\bQ^{-1}\hat\bQ_{i'j'}\hat\bQ^{-1}\bR^T) (\bR\hat\bQ^{-1}\bR^T)^{-1}) )
\end{align}
are evaluated with the variance component estimates from the alternative model.  $\bLambda$ is the inverse of the Hessian matrix and $\Lambda_{ij}$ is its $ij$th element, and $\bR$ is a $d$ by $cd$ matrix with right most $d$ by $d$ matrix a diagonal matrix and all other elements 0. 

Notice that the corrections are marker-specific, and require estimates and partial derivatives from the alternative model. In our experience, without the corrections, the score test is often too conservative, the Wald test is often too anti-conservative, while the likelihood ratio test behaves between the two and has the correct control for type I error.

\subsection{Phenotype Imputation}
The tricks used in our mvLMM algorithms require complete or imputed phenotypes. Although for many studies fully observed phenotypes will be collected for all individuals, it is possible that in some cases, some individuals may have one or more phenotypes missing. To fully harness the information in these situiations and to avoid dropping individuals with partially missing phenotypes, we present a method to impute missing phenotypes before association tests. 

Specifically, we first estimate $\hat\bb$, $\hat \bV_g$ and $\hat \bV_e$ in the null model using individuals with completely observed phenotypes. Afterwards, we impute missing phenotype values using the conditional mean given observed phenotypes and estimated parameters. Denote $n_o$ as the number of observed values, $n_m$ as the number of missing values ($n_o+n_m=nd$), $\by_o$ as a $n_o$ vector of observed values and $\bX_o$ as a $n_o$ by $dc$ matrix of corresponding covariates, $\by_m$ as a $n_m$ vector of missing values and $\bX_m$ as a $n_m$ by $dc$ matrix of corresponding covariates. Under the null mvLMM, $\by=[\by_o^T, \by_m^T]^T$ follows a multivariate normal distribution with covariance matrix $\hat\bH=\bK\otimes \hat\bV_g+\bI\otimes \hat\bV_e$. Let $\hat\bH_{oo}$ be the $n_o$ by $n_o$ sub-matrix of $\bH$ that corresponds to the $n_o$ observed values, and $\hat\bH_{mo}$ be the $n_m$ by $n_o$ sub-matrix of $\hat\bH$ that corresponds to the $n_m$ missing values and $n_o$ observed values. We use the conditional mean of $\by_m$ given $\by_o$ and estimated parameters as an estimate for the missing values
\begin{equation}
\hat\by_m=\bX_m^T\hat\bb+ \hat\bH_{mo}\hat\bH_{oo}^{-1}(\by_o-\bX_o^T\hat\bb).
\end{equation}

\clearpage
\newpage
\section{Supplementary Results}

\setcounter{figure}{0}
\setcounter{table}{0}
\makeatletter 
\renewcommand{\thefigure}{S\@arabic\c@figure} 
\makeatletter 
\renewcommand{\thetable}{S\@arabic\c@table}

\begin{figure}[htb!]
\centering
   \includegraphics[width=150pt, height=150pt, angle=0]{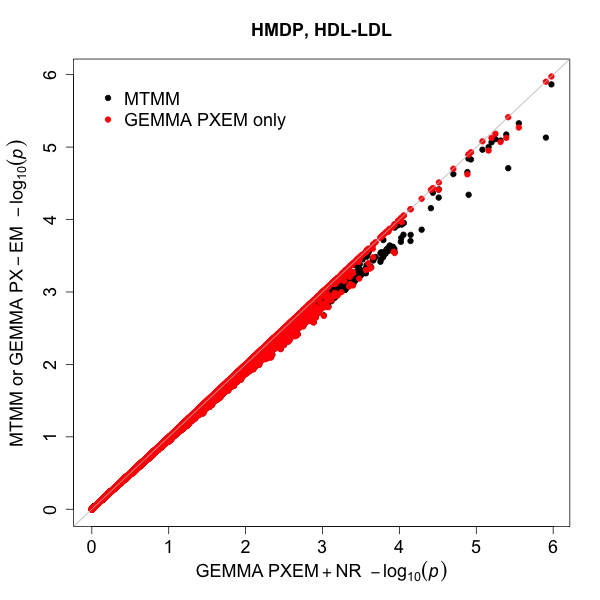}
   \includegraphics[width=150pt, height=150pt, angle=0]{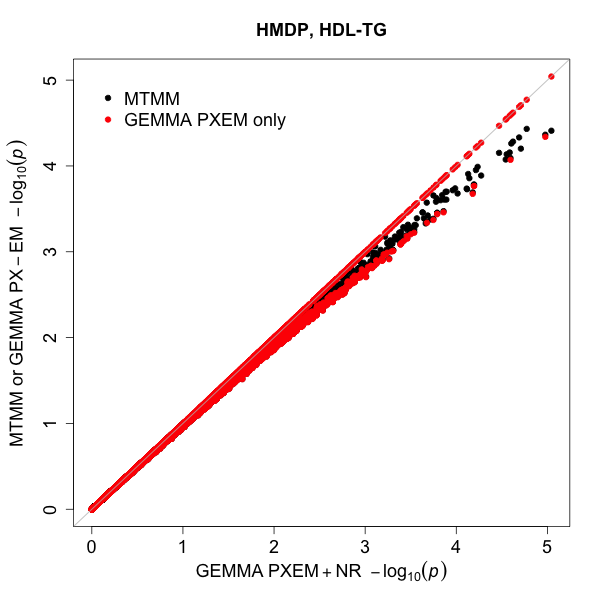}
   \includegraphics[width=150pt, height=150pt, angle=0]{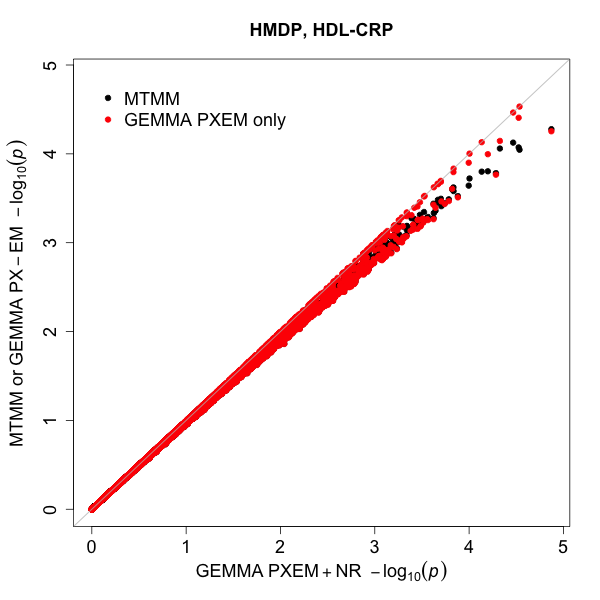}
    \includegraphics[width=150pt, height=150pt, angle=0]{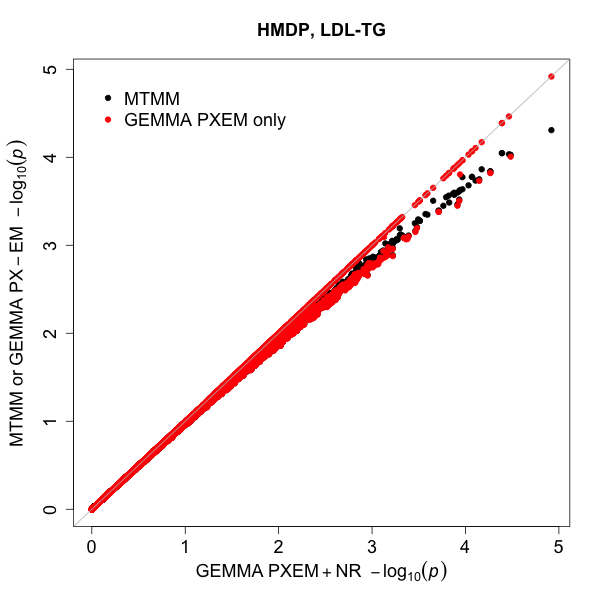}
    \includegraphics[width=150pt, height=150pt, angle=0]{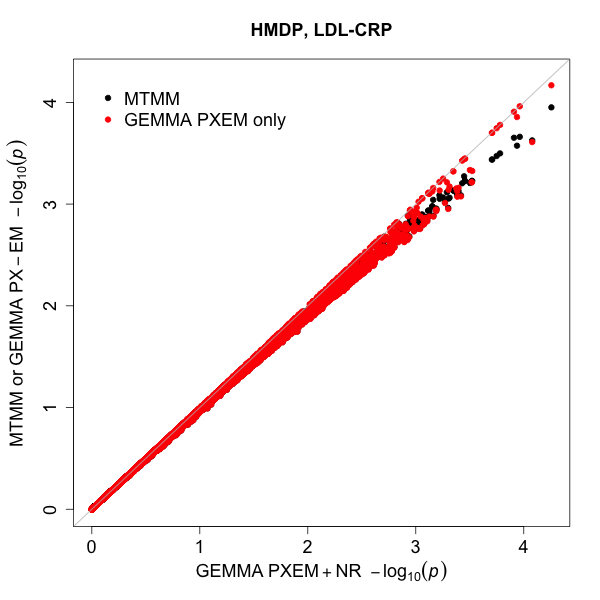}
    \includegraphics[width=150pt, height=150pt, angle=0]{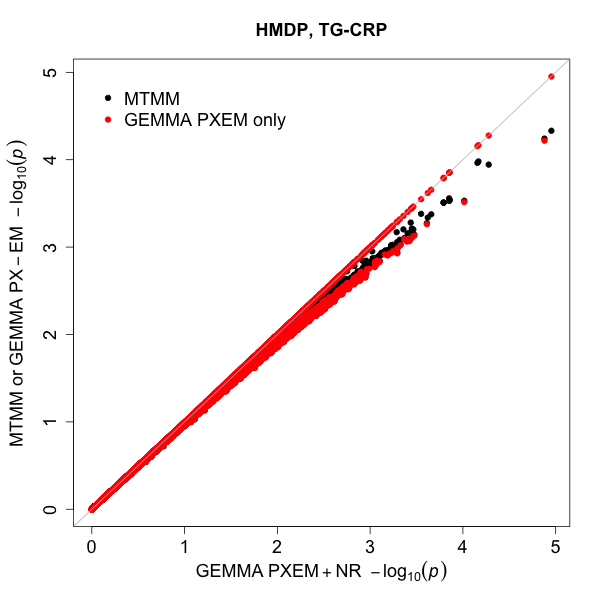}
\caption{Comparison of -$\log_{10}p$ values obtained from GEMMA using both PX-EM and NR, with those from MTMM (black) or those from GEMMA using only PX-EM (red), for all paired traits in the HMDP data set.}
\label{sfig:pxemvsboth}
\end{figure}


\clearpage
\newpage
\begin{figure}[htb!]
\centering
\subfigure[HMDP-based simulations]{
   \includegraphics[width=210pt, height=210pt, angle=0]{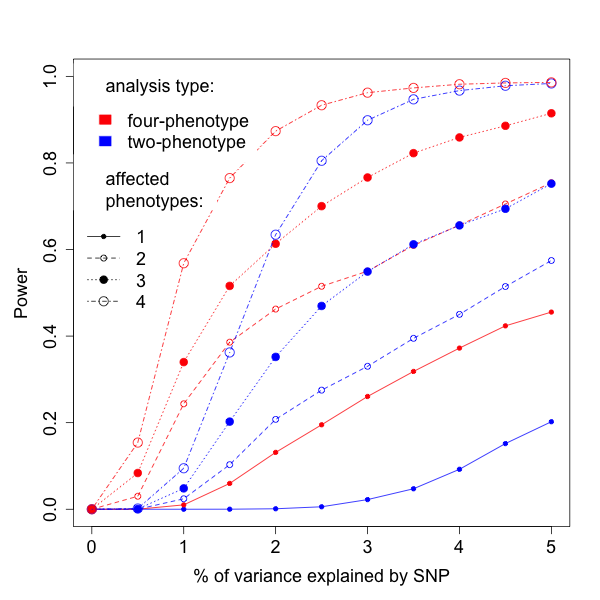}
   \label{sfig:hmdp_4vs2}
 }
 \subfigure[NFBC1966-based simulations]{
   \includegraphics[width=210pt, height=210pt, angle=0]{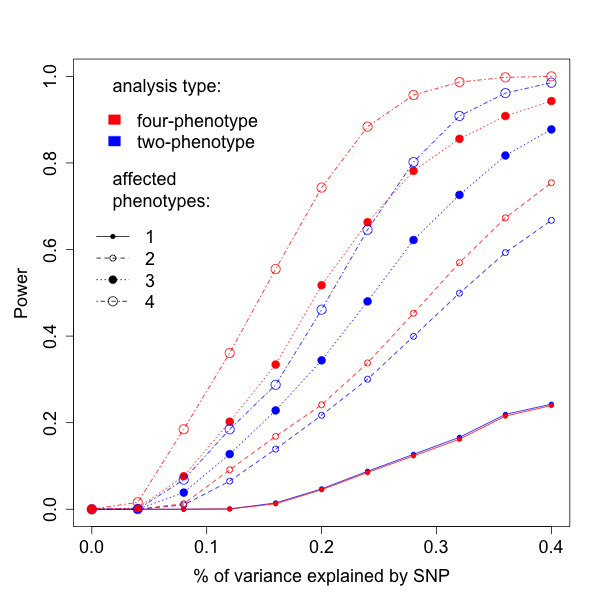}
   \label{sfig:nfbc_4vs2}
 }
 \subfigure[HMDP-based simulations]{
   \includegraphics[width=210pt, height=210pt, angle=0]{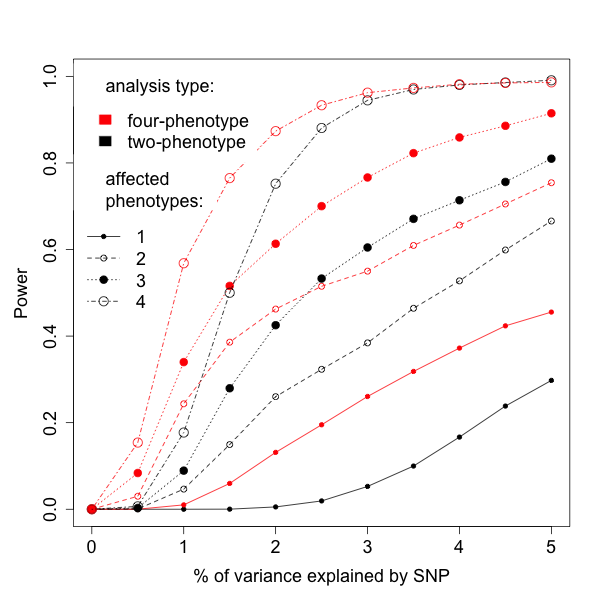}
   \label{sfig:hmdp_4vs2_v2}
 }
 \subfigure[NFBC1966-based simulations]{
   \includegraphics[width=210pt, height=210pt, angle=0]{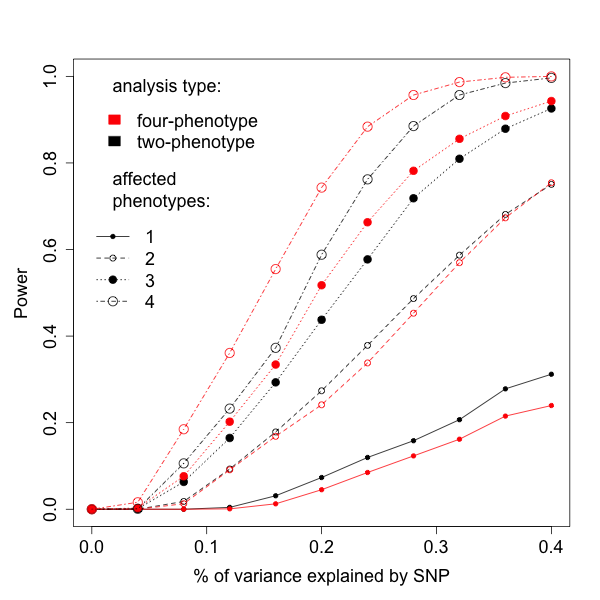}
   \label{sfig:nfbc_4vs2_v2}
 }
\caption{Power comparison between the four-phenotype analysis and the two-phenotype analysis with GEMMA, using simulations based on the HMDP data (a, c) or the NFBC1966 data (b, d). x-axis shows the proportion of phenotypic variance in the affected traits explained (PVE) by the SNP. Symbol size and line type indicate the number of phenotypes affected by the causal SNP. For four-phenotype analysis (red), the genome-wide significance threshold after Bonferroni correction for the number of SNPs is used ($4.6\times 10^{-7}$ for HMDP-based simulations and $1.6\times 10^{-7}$ for NFBC1966-based simulations). For two-phenotype analysis, either the same genome-wide significance threshold is used (black) (c, d), or a significance threshold further corrected for the six tests performed by the two-trait analysis ($7.6\times 10^{-8}$ for HMDP-based simulations and $2.6\times 10^{-8}$ for NFBC1966-based simulations) is used (blue) (a, b).}
\label{sfig:4vs2}
\end{figure}

\clearpage
\newpage
\begin{figure}[htb!]
\centering
\subfigure[Imputation vs Dropping (HMDP)]{
   \includegraphics[width=210pt, height=210pt, angle=0]{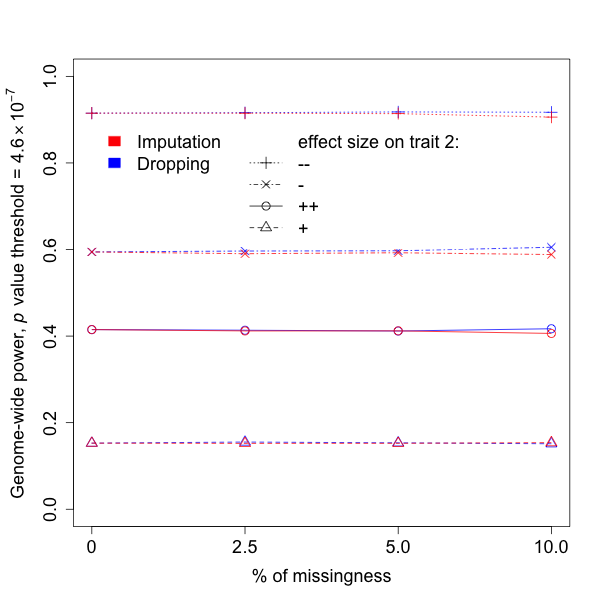}
   \label{sfig:hmdp_impute}
 }
 \subfigure[Imputation vs Dropping (NFBC1966)]{
   \includegraphics[width=210pt, height=210pt, angle=0]{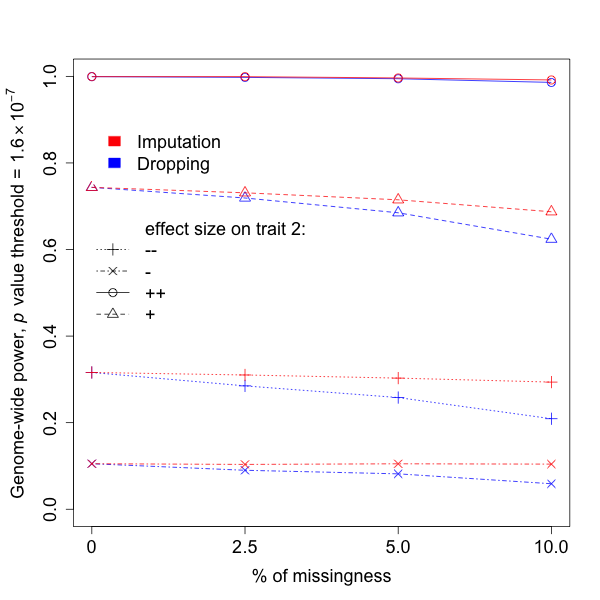}
   \label{sfig:nfbc_impute}
 }
\caption{Power comparison of the two approaches to deal with missing phenotypes in GEMMA, using simulations based on the HMDP data (a) or the NFBC1966 data (b). The first approach use only individuals with fully observed phenotypes (dropping, blue). The second approach imputes phenotypes before association tests (imputation, red). x-axis shows the percentage of individuals having one missing phenotype. the point symbol and line type indicate the SNP effect direction (compared with its effect on the first phenotype) and size (quantified by PVE) on the second phenotype (+: opposite direction, 0.8PVE; $\times$: opposite direction, 0.2PVE; o: same direction, 0.8PVE; $\Delta$: same direction, 0.2PVE). }
\label{sfig:impute}
\end{figure}

\begin{table}
\begin{tiny}
\begin{center}
\caption{List of SNPs that exceed the significance of 0.05 after Bonferroni correction for both the number of SNPs and the number of tests from either the four-phenotype mvLMM analysis or the univariate LMM analysis in the NFBC1966 data set. $\lambda_{GC}$ is the genomic control inflation factor. $p$ values in the four-phenotype analysis below the threshold ($1.6\times 10^{-7}$), and $p$ values in the univariate analysis below the threshold ($3.9\times 10^{-8}$) are in bold font. SNPs that are significant only in the four-phenotype analysis but not in the univariate analysis are highlighted in red, while SNPs that are significant only in the univariate analysis are highlighted in blue. SNPs that are more significant in the four-phenotype analysis than in the univariate analysis (after correcting for the number of tests performed) are highlighted in magenta  (in addition to red). }
\label{tab:p_nfbc_s}
\bigskip
\begin{tabular}{ ccccccc }
\hline
\multirow{3}{*}{SNP}  & \multirow{3}{*}{Position} & mvLMM ($p$ value) & \multicolumn{4}{c}{LMM ($p$ value)} \\
& & Four Traits & HDL & LDL & TG & CRP \\ 
& & ($\lambda_{GC}=0.979$) & ($\lambda_{GC}=0.998$) & ($\lambda_{GC}=1.003$) & ($\lambda_{GC}=1.000$) & ($\lambda_{GC}=0.993$) \\ 
\hline
{\it CELSR2} & chromosome 1 &&&&& \\
\textcolor{magenta}{rs646776} & 109620053 & $\mathbf{1.54\times 10^{-16}}$ & $9.17\times 10^{-12}$ & $\mathbf{1.23\times 10^{-15}}$ & $9.30\times 10^{-1}$ & $6.32\times 10^{-2}$ \\
{\it CRP} & chromosome 1 &&&&& \\
rs1811472 & 157908973 &  $\mathbf{1.19\times 10^{-13}}$ & $7.05\times 10^{-2}$ & $8.81\times 10^{-1}$ & $4.90\times 10^{-1}$ & $\mathbf{1.42\times 10^{-15}}$ \\
rs12093699&157914612&$\mathbf{4.71\times 10^{-12}}$ & $8.75\times 10^{-1}$ & $5.77\times 10^{-1}$ & $8.39\times 10^{-1}$ & $\mathbf{3.39\times 10^{-13}}$ \\
rs2592887&157919563& $\mathbf{8.79\times 10^{-16}}$ & $4.99 \times 10^{-2}$ & $9.34\times 10^{-1}$ & $3.17\times 10^{-1}$ & $\mathbf{5.13\times 10^{-18}}$\\
rs2794520&157945440& $\mathbf{9.29\times 10^{-21}}$ & $2.98\times 10^{-1}$ & $7.34\times 10^{-1}$ & $9.91\times 10^{-1}$ & $\mathbf{2.84\times 10^{-22}}$\\
\textcolor{magenta}{rs11265260} &157966663& $\mathbf{2.27\times 10^{-10}}$ & $5.85\times 10 ^{-2}$ & $1.55\times 10^{-1}$ & $6.90\times 10^{-1}$ & $\mathbf{6.28\times 10^{-11}}$\\
{\it APOB} & chromosome 2 &&&&& \\
\textcolor{red}{rs6728178} & 21047434 & $\mathbf{7.37\times 10^{-9}}$ & $3.46\times 10^{-6}$ & $7.57\times 10^{-7}$ & $3.29\times 10^{-6}$ & $3.62\times 10^{-1}$\\
\textcolor{red}{rs6754295} & 21059688 & $\mathbf{7.99\times 10^{-9}}$ & $3.83\times 10^{-6}$ & $6.27\times 10^{-7}$ & $5.25\times 10^{-6}$ & $3.48\times 10^{-1}$\\
\textcolor{red}{rs676210} & 21085029 & $\mathbf{2.09\times 10^{-8}}$ & $2.74\times 10^{-6}$ & $8.01\times 10^{-6}$ & $1.79\times 10^{-6}$ & $4.65\times 10^{-1}$ \\
rs693&21085700& $\mathbf{9.50\times 10^{-8}}$ & $4.19\times 10 ^{-2}$ & $\mathbf{1.05\times 10^{-9}}$ &$4.00\times 10^{-3}$ & $3.48\times 10^{-1}$\\
\textcolor{red}{rs673548} & 21091049 & $\mathbf{1.36\times 10^{-8}}$ & $2.19\times 10^{-6}$ & $6.71\times 10^{-6}$ & $1.32\times 10^{-6}$ & $4.79\times 10^{-1}$ \\
\textcolor{blue}{rs1429974}&21154275& $6.33\times 10^{-7}$ & $8.02\times 10^{-2}$ & $\mathbf{1.65\times 10^{-8}}$ & $1.84\times 10^{-1}$ & $9.48\times 10^{-1}$\\
rs754524&21165046& $\mathbf{2.44\times 10^{-8}}$ & $9.44\times 10^{-2}$ & $\mathbf{7.82\times 10^{-10}}$ & $1.57\times 10^{-1}$ & $6.10\times 10^{-1}$\\
\textcolor{blue}{rs754523} &21165196& $5.68\times 10^{-7}$ &$8.55\times 10^{-2}$ & $\mathbf{1.55\times 10^{-8}}$ & $1.93\times 10^{-1}$ & $9.66\times 10^{-1}$\\
{\it GCKR} & chromosome 2 &&&&& \\
rs1260326&27584444& $\mathbf{1.16\times 10^{-8}}$ & $1.58\times 10^{-1}$&$1.14\times 10^{-1}$&$\mathbf{2.21\times 10^{-10}}$ & $5.25\times 10^{-2}$\\
rs780094&27594741& $\mathbf{1.02\times 10^{-7}}$ & $3.38\times 10^{-1}$ & $2.63\times 10^{-1}$& $\mathbf{3.50\times 10^{-9}}$ & $1.44\times 10^{-1}$\\
{\it PPP1R3B} & chromosome 8 &&&&& \\
\textcolor{red}{rs983309}&9215142& $\mathbf{6.40\times 10^{-9}}$ & $5.99\times 10^{-5}$ & $2.57\times 10^{-3}$ & $7.57\times 10^{-1}$&$1.55\times 10^{-3}$\\
\textcolor{red}{rs2126259}&9222556& $\mathbf{1.19\times 10^{-9}}$ & $1.67\times 10^{-5}$ & $5.79\times 10^{-4}$ & $4.24\times 10^{-1}$& $5.93\times 10^{-3}$\\
{\it LPL} & chromosome 8 &&&&& \\
\textcolor{magenta}{rs10096633}&19875201& $\mathbf{8.96\times 10^{-10}}$ & $8.23\times 10^{-7}$ & $9.36\times 10^{-1}$ & $\mathbf{1.01\times10^{-8}}$ & $5.22\times 10^{-1}$\\
{\it FADS} & chromosome 11 &&&&& \\
\textcolor{red}{rs174537}&61309256& $\mathbf{5.55\times 10^{-8}}$ & $3.41\times 10^{-2}$ & $3.76\times 10^{-6}$ & $2.45\times 10^{-2}$ & $8.44\times 10^{-1}$\\
\textcolor{red}{rs102275}&61314379& $\mathbf{3.25\times 10^{-8}}$ & $2.22\times 10^{-2}$ & $3.06\times 10^{-6}$ & $2.27\times 10^{-2}$ & $9.05\times 10^{-1}$\\
\textcolor{red}{rs174546}&61326406& $\mathbf{2.80\times 10^{-8}}$ & $3.99\times 10^{-2}$ & $2.25\times 10^{-6}$ & $2.03\times 10^{-2}$ & $9.04\times 10^{-1}$\\
\textcolor{red}{rs174556}&61337211& $\mathbf{6.88\times 10^{-8}}$ & $1.67\times 10^{-1}$ & $7.02\times 10^{-7}$ & $5.44\times 10^{-2}$& $9.89\times 10^{-1}$\\
\textcolor{red}{rs1535}&61354548& $\mathbf{9.40\times 10^{-8}}$ & $4.76\times 10^{-2}$ & $4.11\times 10^{-6}$ & $2.77\times 10^{-2}$ & $8.71\times 10^{-1}$\\
{\it HNF1A} & chromosome 12 &&&&& \\
rs2650000&119873345& $\mathbf{1.48\times 10^{-10}}$ & $2.16\times 10^{-1}$ & $9.36\times 10^{-1}$ & $6.48\times 10^{-1}$ & $\mathbf{1.44\times 10^{-12}}$\\
rs7953249&119888107& $\mathbf{2.21\times 10^{-10}}$ & $1.43\times 10^{-1}$ & $9.16\times 10^{-1}$ & $6.10\times 10^{-1}$ & $\mathbf{2.38\times 10^{-12}}$\\
rs1169300&119915608& $\mathbf{5.29\times 10^{-8}}$ & $6.76\times 10^{-1}$ & $4.05\times 10^{-1}$ & $5.27\times 10^{-1}$ & $\mathbf{4.28\times 10^{-9}}$\\
rs2464196&119919810& $\mathbf{5.82\times 10^{-8}}$ & $5.95\times 10^{-1}$ & $5.44\times 10^{-1}$ &$5.74\times 10^{-1}$ & $\mathbf{3.58\times 10^{-9}}$\\
rs735396&119923227& $\mathbf{1.19\times 10^{-7}}$ & $7.17\times 10^{-1}$ & $3.22\times 10^{-1}$ & $3.18\times 10^{-1}$ & $\mathbf{2.24\times 10^{-8}}$\\
{\it LIPC} & chromosome 15 &&&&& \\
\textcolor{red}{rs166358} & 56468097 & $\mathbf{3.66\times 10^{-10}}$ & $8.57\times 10^{-8}$ & $3.80\times 10^{-2}$ & $2.29\times 10^{-1}$ & $5.43\times 10^{-1}$\\
\textcolor{magenta}{rs1532085} &56470658& $\mathbf{2.52\times 10^{-16}}$ & $\mathbf{8.33\times 10^{-13}}$ & $3.46\times 10^{-1}$ & $6.05\times 10^{-2}$ & $6.21\times 10^{-1}$\\
\textcolor{magenta}{rs415799} &56478046& $\mathbf{7.47\times 10^{-10}}$ & $\mathbf{2.32\times 10^{-8}}$ & $5.72\times 10^{-1}$ & $1.13\times 10^{-1}$ & $9.06\times 10^{-1}$\\
\textcolor{red}{rs16940213} & 56482629 & $\mathbf{5.51\times 10^{-8}}$ & $3.19\times 10^{-6}$ & $2.67\times 10^{-1}$ & $1.31\times 10^{-1}$ & $4.29\times 10^{-1}$\\
\textcolor{red}{rs473224}&56524633& $\mathbf{6.54\times 10^{-9}}$ & $1.57\times 10^{-3}$ & $5.30\times 10^{-1}$ & $7.64\times 10^{-5}$ & $1.73\times 10^{-1}$\\
\textcolor{red}{rs261336}&56529710& $\mathbf{2.29\times 10^{-9}}$ & $6.51\times 10^{-4}$ & $2.30\times 10^{-1}$ & $7.08\times 10^{-5}$ & $3.00\times 10^{-1}$\\
{\it CETP} & chromosome 16 &&&&& \\
\textcolor{magenta}{rs9989419}&55542640& $\mathbf{1.88\times 10^{-9}}$ & $\mathbf{5.79\times 10^{-10}}$ & $7.63\times 10^{-1}$ & $9.62\times 10^{-1}$ & $6.12\times 10^{-1}$\\
\textcolor{magenta}{rs3764261}&55550825& $\mathbf{1.85\times 10^{-38}}$ & $\mathbf{6.56\times 10^{-37}}$ & $7.09\times 10^{-2}$ & $2.52\times 10^{-1}$ & $2.06\times 10^{-1}$\\
rs1532624&55562980& $\mathbf{1.30\times 10^{-26}}$ & $\mathbf{3.15\times 10^{-27}}$ & $1.64\times 10^{-1}$ & $1.20\times 10^{-1}$ & $1.05\times 10^{-1}$\\
rs7499892&55564091& $\mathbf{4.01\times 10^{-22}}$ & $\mathbf{9.93\times 10^{-20}}$ & $8.36\times 10^{-1}$ & $4.51\times 10^{-1}$ & $8.09\times 10^{-1}$\\
{\it LCAT} & chromosome 16 &&&&& \\
rs255049 & 66570972 & $\mathbf{5.34\times 10^{-8}}$ & $\mathbf{6.86\times 10^{-9}}$ & $1.44\times 10^{-1}$ & $1.60\times 10^{-1}$ & $7.50\times 10^{-1}$\\
rs255052 & 66582496 & $\mathbf{9.72\times 10^{-8}}$ & $\mathbf{5.98\times 10^{-9}}$ & $2.01\times 10^{-1}$ & $2.36\times 10^{-1}$ & $7.02\times 10^{-1}$\\
{\it LDLR} & chromosome 19 &&&&& \\
\textcolor{blue}{rs11668477} &11056030& $1.09\times 10^{-6}$ & $5.95\times 10^{-2}$ & $\mathbf{2.04\times 10^{-8}}$ & $1.66\times 10^{-2}$ & $7.18\times 10^{-1}$\\
{\it APO cluster} & chromosome 19 &&&&& \\
\textcolor{magenta}{rs157580}&50087106& $\mathbf{1.51\times 10^{-8}}$ & $9.82\times 10^{-3}$& $\mathbf{1.06\times 10^{-8}}$ & $1.18\times 10^{-3}$& $2.42\times 10^{-1}$\\
\textcolor{red}{rs2075650}&50087459& $\mathbf{1.21\times 10^{-12}}$ & $4.95\times 10^{-2}$ & $8.54\times 10^{-6}$ & $2.20\times 10^{-5}$ & $1.01\times 10^{-5}$ \\
{\it HNF4A} & chromosome 20 &&&&& \\
rs1800961 & 42475778& $\mathbf{1.21\times 10^{-7}}$ & $\mathbf{1.15\times 10^{-8}}$ & $3.78\times 10^{-1}$ & $4.59\times 10^{-3}$ & $4.99\times 10^{-1}$\\
\hline
\end{tabular}
\end{center}
\end{tiny}
\end{table}

\begin{table}
\begin{tiny}
\begin{center}
\caption{List of SNPs that exceed the significance of 0.05 after Bonferroni correction for both the number of SNPs and the number of tests from either the four-phenotype mvLMM analysis or the two-phenotype mvLMM analysis in the NFBC1966 data set. $\lambda_{GC}$ is the genomic control inflation factor. $p$ values in the four-phenotype analysis below the threshold ($1.6\times 10^{-7}$), and $p$ values in the two-phenotype analysis below the threshold ($2.6\times 10^{-8}$) are in bold font. SNPs that are significant only in the four-phenotype analysis but not in the two-phenotype analysis are highlighted in red. No SNP is significant only in the two-phenotype analysis. SNPs that are more significant in the four-phenotype analysis than in the two-phenotype analysis (after correcting for the number of tests performed) are highlighted in magenta (in addition to red).}
\label{tab:p_nfbc_m}
\bigskip
\begin{tabular}{ ccccccccc }
\hline
\multirow{3}{*}{SNP}  & \multirow{3}{*}{Position} & \multicolumn{7}{c}{mvLMM ($p$ value)} \\
& & Four Traits & HDL-LDL & HDL-TG & HDL-CRP & LDL-TG & LDL-CRP & TG-CRP \\ 
& &  ($\lambda_{GC}=0.979$) & ($\lambda_{GC}=0.989$) & ($\lambda_{GC}=0.991$) & ($\lambda_{GC}=0.992$) & ($\lambda_{GC}=0.993$) & ($\lambda_{GC}=0.998$) & ($\lambda_{GC}=0.994$) \\ 
\hline
{\it CELSR2} & chromosome 1 &&&&&&& \\
\textcolor{magenta}{rs646776} &109620053& $\mathbf{1.54\times 10^{-16}}$ & $\mathbf{1.1\times 10^{-14}}$ & $1.98\times 10^{-1}$ & $1.92\times 10^{-2}$ & $\mathbf{3.06\times 10^{-16}}$ & $\mathbf{3.84\times 10^{-16}}$ & $1.61\times 10^{-1}$ \\
{\it CRP} & chromosome 1 &&&&&&& \\
rs1811472&157908973& $\mathbf{1.19\times 10^{-13}} $& $1.82\times 10^{-1}$ & $2.04\times 10^{-1}$ & $\mathbf{9.17\times 10^{-15}}$ & $7.18\times 10^{-1}$ & $\mathbf{9.23\times 10^{-15}}$ & $\mathbf{7.87\times 10^{-15}}$ \\
rs12093699&157914612& $\mathbf{4.71\times 10^{-12}}$ & $8.28\times 10^{-1}$ & $9.70\times 10^{-1}$ & $\mathbf{3.82\times 10^{-13}}$ & $7.99\times 10^{-1}$ & $\mathbf{3.17\times 10^{-12}}$ & $\mathbf{5.40\times 10^{-13}}$ \\
rs2592887&157919563& $\mathbf{8.79\times 10^{-16}}$ & $1.45\times 10^{-1}$ & $1.49\times 10^{-1}$ & $\mathbf{4.24\times 10^{-17}}$ & $5.80\times 10^{-1}$ & $\mathbf{4.12\times 10^{-17}}$ & $\mathbf{3.97\times 10^{-17}}$\\
rs2794520&157945440& $\mathbf{9.29\times 10^{-21}}$ & $5.20\times 10^{-1}$ & $5.34\times 10^{-1}$ & $\mathbf{1.77\times 10^{-21}}$ & $9.31\times 10^{-1}$ & $\mathbf{1.47\times 10^{-21}}$ & $\mathbf{2.82\times 10^{-22}}$\\
\textcolor{magenta}{rs11265260}&157966663& $\mathbf{2.27\times 10^{-10}}$ & $3.98\times 10^{-2}$ & $7.62\times 10^{-2}$ & $\mathbf{5.11\times 10^{-10}}$ & $3.68\times 10^{-1}$ & $\mathbf{5.94\times 10^{-11}}$ & $\mathbf{6.07\times 10^{-11}}$ \\
{\it APOB} & chromosome 2 &&&&&&& \\
\textcolor{magenta}{rs6728178}&21047434& $\mathbf{7.37\times 10^{-9}}$ & $\mathbf{1.48\times 10^{-9}}$ & $2.03\times 10^{-7}$ & $2.63\times 10^{-5}$ & $3.02\times 10^{-8}$ & $4.92\times 10^{-6}$ & $2.23\times 10^{-5}$ \\
\textcolor{magenta}{rs6754295}&21059688& $\mathbf{7.99\times 10^{-9}}$ & $\mathbf{1.37\times 10^{-9}}$ & $2.98\times 10^{-7}$ & $2.78\times 10^{-5}$ & $3.61\times 10^{-8}$ & $4.01\times 10^{-6}$ & $3.40\times 10^{-5}$ \\
\textcolor{magenta}{rs676210}&21085029& $\mathbf{2.09\times 10^{-8}} $ & $\mathbf{9.04\times 10^{-9}}$ & $1.11\times 10^{-7}$ & $2.03\times 10^{-5}$ & $1.07\times 10^{-7}$ & $4.91\times 10^{-5}$ & $1.18\times 10^{-5}$ \\
rs693&21085700& $\mathbf{9.50\times 10^{-8}}$ & $\mathbf{3.95\times 10^{-9}}$ & $9.83\times 10^{-3}$ & $1.26\times 10^{-1}$ & $\mathbf{5.94\times 10^{-9}}$ & $\mathbf{9.09\times 10^{-9}}$ & $1.81\times 10^{-2}$\\
\textcolor{magenta}{rs673548}&21091049& $\mathbf{1.36\times 10^{-8}}$ & $\mathbf{6.45\times 10^{-9}}$ & $7.78\times 10^{-8}$ & $1.64\times 10^{-5}$ & $7.52\times 10^{-8}$ & $4.18\times 10^{-5}$ & $8.73\times 10^{-6}$\\
rs754524&21165046& $\mathbf{2.44\times 10^{-8}}$ & $\mathbf{4.15\times 10^{-9}}$ & $1.79\times 10^{-1}$ & $1.63\times 10^{-1}$ & $\mathbf{4.90\times 10^{-9}}$ & $\mathbf{3.01\times 10^{-9}}$ & $2.46\times 10^{-1}$ \\
{\it GCKR} & chromosome 2 &&&&&&& \\
rs1260326&27584444&$\mathbf{1.16\times 10^{-8}}$ & $1.38\times 10^{-1}$ & $\mathbf{9.00\times 10^{-10}}$ & $8.54\times 10^{-2}$ & $\mathbf{1.51\times 10^{-9}}$ & $5.77\times 10^{-2}$ & $\mathbf{1.36\times 10^{-9}}$ \\
rs780094&27594741& $\mathbf{1.02\times 10^{-7}}$ & $3.83\times 10^{-1}$ & $\mathbf{9.08\times 10^{-9}}$ & $2.61\times 10^{-1}$ & $\mathbf{1.82\times 10^{-8}}$ & $2.11\times 10^{-1}$ & $\mathbf{2.23\times 10^{-8}}$\\
{\it PPP1R3B} & chromosome 8 &&&&&&& \\
\textcolor{red}{rs983309} &9215142& $\mathbf{6.40\times 10^{-9}}$ & $5.04\times 10^{-7}$ & $1.24\times 10^{-4}$ & $7.54\times 10^{-8}$ & $3.98\times 10^{-3}$ & $1.71\times 10^{-4}$ & $3.30\times 10^{-3}$\\
\textcolor{magenta}{rs2126259}&9222556& $\mathbf{1.19\times 10^{-9}}$ & $\mathbf{2.41\times 10^{-8}}$ & $6.04\times 10^{-5}$ & $1.02\times 10^{-7}$ & $3.11\times 10^{-4}$ & $1.46\times 10^{-4}$ & $6.95\times 10^{-3}$\\
{\it LPL} & chromosome 8 &&&&&&& \\
\textcolor{magenta}{rs10096633}&19875201& $\mathbf{8.96\times 10^{-10}}$ & $4.38\times 10^{-6}$ & $\mathbf{1.10\times 10^{-9}}$ & $2.53\times 10^{-6}$ & $\mathbf{1.20\times 10^{-8}}$ & $8.12\times 10^{-1}$ & $3.01\times 10^{-8}$\\
{\it FADS} & chromosome 11 &&&&&&& \\
rs174537&61309256& $\mathbf{5.55\times 10^{-8}}$ & $5.10\times 10^{-7}$ & $3.43\times 10^{-2}$ & $8.73\times 10^{-2}$ & $\mathbf{8.47\times 10^{-9}}$ & $2.32\times 10^{-5}$ & $5.96\times 10^{-2}$ \\
\textcolor{magenta}{rs102275}&61314379& $\mathbf{3.25\times 10^{-8}}$ & $2.54\times 10^{-7}$ & $2.51\times 10^{-2}$ & $6.23\times 10^{-2}$ & $\mathbf{5.90\times 10^{-9}}$ & $1.85\times 10^{-5}$ & $5.91\times 10^{-2}$\\
rs174546&61326406& $\mathbf{2.80\times 10^{-8}}$ & $3.6\times 10^{-7}$ & $3.32\times 10^{-2}$ & $1.06\times 10^{-1}$ & $\mathbf{3.35\times 10^{-9}}$ & $1.38\times 10^{-5}$ & $5.34\times 10^{-2}$\\
rs174556&61337211& $\mathbf{6.88\times 10^{-8}}$ & $5.58\times 10^{-7}$ & $1.32\times 10^{-1}$ & $3.84\times 10^{-1}$ & $\mathbf{4.08\times 10^{-9}}$ & $4.04\times 10^{-6}$ & $1.45\times 10^{-1}$ \\
rs1535&61354548& $\mathbf{9.40\times 10^{-8}}$ & $8.18\times 10^{-7}$ & $4.48\times 10^{-2}$ & $1.21\times 10^{-1}$ & $\mathbf{1.18\times 10^{-8}}$ & $2.50\times 10^{-5}$ & $6.91\times 10^{-2}$\\
{\it HNF1A} & chromosome 12 &&&&&&& \\
rs2650000&119873345& $\mathbf{1.48\times 10^{-10}}$ & $4.54\times 10^{-1}$ & $4.53\times 10^{-1}$ & $\mathbf{1.67\times 10^{-11}}$ & $8.76\times 10^{-1}$ & $\mathbf{1.02\times 10^{-11}}$ & $\mathbf{4.90\times 10^{-12}}$\\
rs7953249&119888107& $\mathbf{2.21\times 10^{-10}}$ & $3.31\times 10^{-1}$ & $3.35\times 10^{-1}$ & $\mathbf{2.45\times 10^{-11}}$ & $8.41\times 10^{-1}$ & $\mathbf{1.67\times 10^{-11}}$ & $\mathbf{9.40\times 10^{-12}}$\\
rs1169300&119915608& $\mathbf{5.29\times 10^{-8}}$ & $6.16\times 10^{-1}$ & $6.20\times 10^{-1}$ & $\mathbf{2.55\times 10^{-8}}$ & $6.81\times 10^{-1}$ & $\mathbf{1.26\times 10^{-8}}$ & $\mathbf{3.13\times 10^{-9}}$ \\
rs2464196&119919810& $\mathbf{5.82\times 10^{-8}}$ & $6.90\times 10^{-1}$ & $5.99\times 10^{-1}$ & $\mathbf{2.31\times 10^{-8}}$ & $7.92\times 10^{-1}$ & $\mathbf{1.44\times 10^{-8}}$ & $\mathbf{3.05\times 10^{-9}}$\\
rs735396&119923227& $\mathbf{1.19\times 10^{-7}}$ & $5.45\times 10^{-1}$ & $4.23\times 10^{-1}$ & $1.17\times 10^{-7}$ & $4.96\times 10^{-1}$ & $5.24\times 10^{-8}$ & $\mathbf{8.3\times 10^{-9}}$\\
{\it LIPC} & chromosome 15 &&&&&&& \\
\textcolor{magenta}{rs166358}&56468097& $\mathbf{3.66\times 10^{-10}}$ & $2.34\times 10^{-7}$ & $\mathbf{8.28\times 10^{-10}}$ & $1.39\times 10^{-7}$ & $1.58\times 10^{-2}$ & $8.50\times 10^{-2}$ & $4.45\times 10^{-1}$\\
rs1532085&56470658& $\mathbf{2.52\times 10^{-16}}$ & $\mathbf{1.21\times 10^{-12}}$ & $\mathbf{1.06\times 10^{-17}}$ & $\mathbf{1.21\times 10^{-12}}$ & $1.67\times 10^{-1}$ & $5.87\times 10^{-1}$ & $1.63\times 10^{-1}$\\
rs415799&56478046& $\mathbf{7.47\times 10^{-10}}$ & $7.02\times 10^{-8}$ & $\mathbf{2.87\times 10^{-11}}$ & $1.03\times 10^{-7}$ & $2.90\times 10^{-1}$ & $8.39\times 10^{-1}$ & $2.39\times 10^{-1}$\\
\textcolor{magenta}{rs16940213}&56482629& $\mathbf{5.51\times 10^{-8}}$ & $1.71\times 10^{-5}$ & $\mathbf{2.37\times 10^{-8}}$ & $3.84\times 10^{-6}$ & $7.46\times 10^{-2}$ & $3.62\times 10^{-1}$ & $2.85\times 10^{-1}$\\
rs473224&56524633& $\mathbf{6.54\times 10^{-9}}$ &$3.80\times 10^{-3}$ & $\mathbf{7.00\times 10^{-10}}$ & $9.57\times 10^{-4}$ & $3.09\times 10^{-4}$ & $3.47\times 10^{-1}$ & $2.80\times 10^{-4}$\\
rs261336&56529710& $\mathbf{2.29\times 10^{-9}}$ & $7.29\times 10^{-4}$ & $\mathbf{1.33\times 10^{-10}}$ & $6.20\times 10^{-4}$ & $3.68\times 10^{-4}$ & $3.19\times 10^{-1}$ & $3.35\times 10^{-4}$\\
{\it CETP} & chromosome 16 &&&&&&& \\
rs9989419&55542640& $\mathbf{1.88\times 10^{-9}}$ & $\mathbf{4.05\times 10^{-9}}$ & $\mathbf{9.67\times 10^{-11}}$ & $\mathbf{2.94\times 10^{-9}}$ & $9.21\times 10^{-1}$ & $8.41\times 10^{-1}$ & $8.43\times 10^{-1}$ \\
rs3764261&55550825& $\mathbf{1.85\times 10^{-38}}$ & $\mathbf{1.06\times 10^{-35}}$ & $\mathbf{1.19\times 10^{-39}}$ & $\mathbf{2.19\times 10^{-36}}$ & $1.64\times 10^{-1}$ & $1.07\times 10^{-1}$ & $3.02\times 10^{-1}$\\
rs1532624&55562980& $\mathbf{1.30\times 10^{-26}}$ & $\mathbf{4.42\times 10^{-26}}$ & $\mathbf{4.79\times 10^{-28}}$ & $\mathbf{2.04\times 10^{-26}}$ & $1.89\times 10^{-1}$ & $1.25\times 10^{-1}$ & $1.22\times 10^{-1}$\\
rs7499892&55564091& $\mathbf{4.01\times 10^{-22}}$ & $\mathbf{3.88\times 10^{-19}}$ & $\mathbf{1.13\times 10^{-23}}$ & $\mathbf{2.25\times 10^{-19}}$ & $7.52\times 10^{-1}$ & $9.44\times 10^{-1}$ & $6.62\times 10^{-1}$\\
{\it LCAT} & chromosome 16 &&&&&&& \\
rs255049&66570972& $\mathbf{5.34\times 10^{-8}}$ & $\mathbf{4.07\times 10^{-9}}$ & $3.25\times 10^{-8}$ & $2.85\times 10^{-8}$ & $4.75\times 10^{-2}$ & $3.29\times 10^{-1}$ & $3.32\times 10^{-1}$\\
rs255052&66582496& $\mathbf{9.72\times 10^{-8}}$ & $\mathbf{5.24\times 10^{-9}}$ & $\mathbf{2.10\times 10^{-8}}$ & $4.58\times 10^{-8}$ & $1.06\times 10^{-1}$ & $3.81\times 10^{-1}$ & $5.12\times 10^{-1}$\\
{\it APO cluster} & chromosome 19 &&&&&&& \\
\textcolor{magenta}{rs157580}&50087106& $\mathbf{1.51\times 10^{-8}}$ & $\mathbf{1.42\times 10^{-8}}$ & $1.95\times 10^{-3}$ & $7.83\times 10^{-3}$ & $\mathbf{2.57\times 10^{-8}}$ & $\mathbf{1.60\times 10^{-8}}$ & $7.35\times 10^{-4}$\\
\textcolor{magenta}{rs2075650}&50087459& $\mathbf{1.21\times 10^{-12}}$ & $1.93\times 10^{-5}$ & $1.17\times 10^{-4}$ & $7.21\times 10^{-7}$ & $6.24\times 10^{-7}$ & $\mathbf{2.88\times 10^{-10}}$ & $\mathbf{2.27\times 10^{-11}}$\\
{\it HNF4A} & chromosome 20 &&&&&&& \\
rs1800961&42475778& $\mathbf{1.21\times 10^{-7}}$ & $8.41\times 10^{-8}$ & $7.61\times 10^{-8}$ & $\mathbf{1.05\times 10^{-8}}$ & $1.81\times 10^{-2}$ & $5.05\times 10^{-1}$ & $6.03\times 10^{-3}$\\
\hline
\end{tabular}
\end{center}
\end{tiny}
\end{table}

\clearpage
\newpage
\bibliographystyle{unsrt}
\bibliography{../../BIB/lmm,../../BIB/bslmm,../../BIB/mvlmm}

\end{document}